\documentclass[pra,aps,twocolumn,amsmath,amssymb,superscriptaddress]{revtex4-1}
\usepackage{amstext}
\usepackage{amssymb}
\usepackage{graphicx}
\usepackage{color}
\usepackage{amssymb,amsmath}
\usepackage{color}
\usepackage{enumitem}

\def\beq{\begin{equation}}
\def\eeq{\end{equation}}
\def\beal{\begin{align}}
\def\ealg{\end{align}}
\def\bearr{\begin{equation}\begin{array}{ll}}
\def\enarr{\end{array}\end{equation}}

\begin{document}


\title{Distinguishability, degeneracy and correlations in three harmonically trapped bosons in one-dimension}

\author{M. A. Garc\'{\i}a-March} 
\affiliation{Departament d'Estructura i Constituents de la Mat\`eria, Univ. de Barcelona, 08028 Barcelona, Spain}
\author{B. Juli\'a-D\'iaz}
\affiliation{Departament d'Estructura i Constituents de la Mat\`eria, Univ. de Barcelona, 08028 Barcelona, Spain}
\affiliation{ICFO-Institut de Ci\`encies Fot\`oniques, Parc Mediterrani de la Tecnologia, 08860 Barcelona, Spain}
\author{G.~E. Astrakharchik}
\affiliation{Departament de F\'\i sica i Enginyeria Nuclear, Campus Nord B4, Universitat Polit\`ecnica de Catalunya, E-08034 Barcelona, Spain}
\author{J. Boronat} 
\affiliation{Departament de F\'\i sica i Enginyeria Nuclear, Campus Nord B4, Universitat Polit\`ecnica de Catalunya, E-08034 Barcelona, Spain}
\author{A. Polls}
\affiliation{Departament d'Estructura i Constituents de la Mat\`eria, Univ. de Barcelona, 08028 Barcelona, Spain}

\begin{abstract}
We  study a system of two bosons of one species and a third boson of a second species in a one-dimensional parabolic trap at zero temperature. We assume contact repulsive inter- and intra-species interactions. By means of an exact diagonalization method we calculate the ground and excited states for the whole range of interactions. We use discrete group theory to classify the eigenstates according to the symmetry of the interaction potential. We also propose and validate analytical ansatzs gaining physical insight over the numerically obtained wavefunctions. We show that, for both approaches, it is crucial to take into account that the distinguishability of the third atom implies the absence of any restriction over the wavefunction when interchanging this boson with any of the other two. 
We find that there are degeneracies in the spectra in some limiting regimes, that is, when the inter-species and/or the intra-species interactions tend to infinity. This is in contrast with the three-identical boson system, where no degeneracy occurs in these limits. We show that, when tuning both types of interactions  through a protocol that keeps them equal while they are increased towards infinity, the systems's ground state resembles that  of three indistinguishable bosons. Contrarily, the systems's ground state is different from that of three-identical bosons when both types of interactions are increased towards infinity through protocols that do not restrict them to be equal. We study the coherence and correlations of the system as the interactions are tuned through different protocols, which permit to built up different correlations in the system and lead to different spatial distributions of the three atoms.
\end{abstract}

\maketitle

\section{Introduction}

The renewed interest in the theoretical research on systems of a few trapped ultracold bosons or fermions is strongly related to the recent experimental achievements in this direction~\cite{He:10,Serwane:11,Zurn:12,Wenz:13,Bourgain:13,Pagano:14}. In the pioneering works in this topic, the energy spectra of two trapped atoms was obtained analytically~\cite{Busch:98,Idziaszek:06}. This research was  recently  followed by the study  of  the energy spectra of three atoms in a trap in three and reduced dimensions~\cite{Kestner:07,Liu:10,Gharashi:12,Sowinski:13,DAmico:13,Blume:14}. Systems of few atoms permit to study microscopically many interesting phenomena, like the Bardeen-Cooper-Schrieffer to Bose-Einstein-condensate (BCS-BEC) crossover~\cite{Blume:12}. Particularly, one would naturally study strongly correlated regimes, which for bosons trapped in one dimension were described to be a  Tonks-Girardeau (TG) gas~\cite{Girardeau:60, Girardeau:01,Gangardt:03}. The experimental realization of the TG gas~\cite{
Paredes:04,
Kinoshita:04}, paved the way for the theoretical study of new 
phenomena in these systems, like  non-equilibrium dynamics~\cite{cradle}, interferometry~\cite{Fogarty:2013}, or the breathing of an impurity within an interacting gas of bosons~\cite{fazio}.   

Mixtures of a few bosons constitute a very exciting raw material to study strongly correlated regimes. Indeed, in the strongly interacting limit they have features in common with the TG gas. For example, in certain interacting limits, their ground-state wavefunction can be obtained analytically and it is similar to a TG gas for  both components~\cite{Girardeau:07,Deuretzbacher:08}. The intra- and inter-species interactions, which describe all the interaction processes in these mixtures, can be controlled experimentally by means of  Feshbach and confinement induced resonances~\cite{Olshanii1998,Papp:08,Thalhammer:08}. By playing with both the intra- and inter-species interactions one can explore different physical phenomena, like phase separation on small atom mixtures~\cite{Cazalilla:03, Alon:06,Mishra:07,Kleine:08,Garcia-march:12}. Other relevant phenomena are the presence of a composite fermionized gas~\cite{Zollner:08a,Hao:09,Hao:09b}, quantum magnetism~\cite{Deuretzbacher:14,Dehkharghani:14}, or a 
crossover between 
composite fermionization and phase-separation~\cite{Garcia-March:13,Garcia-March:14}. Also, these small number bosonic mixtures allow for the study of dynamical phenomena, like the tunneling of one species through the barrier formed by the other species~\cite{Pflanzer:2009,Pflanzer:2010}
or  the dynamical emergence of  orthogonality catastrophe~\cite{Campbell:14}.

This latter, extremely appealing phenomena, occurs in a system  of two atoms in one species and a third atom in a second species. Such a small system has attracted recently a great interest. For example, how to relate spatial symmetries with the energy spectra in this system, and the differences with a system of three indistinguishable atoms were discussed in Refs.~\cite{Harshman:12,Harshman:14}. Analytical expressions for the wavefunction, which are exact  for some limiting values of the inter- and intra-species interactions, have been also recently obtained, together with the relationship of the system with  anyonic particles, which show fractional statistics~\cite{Zinner:13,Zinner:14}. In this paper we discuss how the distinguishability of the third atom with respect to the other  two identical bosons induces degeneracies for some limiting values of the intra- and inter-species interactions. We use a many-mode  exact diagonalization  method, as the one discussed in Refs.~\cite{Garcia-March:13,Garcia-March:14},  to obtain the 
whole spectra for different values of the interactions together with the corresponding wavefunctions. We propose analytical ansatzs for the wavefunction of different states with straightforward physical interpretations in each limit. We show that these anstatzs approximate accurately the actual wavefunctions  in those limits. We also show that they are in agreement with the discrete group theory results of  Refs.~\cite{Harshman:12,Harshman:14}. We study the coherence and correlations of the system as the interactions are tuned. Since there are two different types of interaction in the system, there are different protocols which one can use to tune the interactions. We demonstrate that these protocols can be used to build different  correlations in the system. For example, if both types of interactions are equal, the wavefunction of the ground state corresponds to that of a gas of three indistinguishable bosons for all values of the interactions. Therefore,  they behave as a three-
boson TG gas  when the 
interactions are large. On the contrary, following other protocols, for example increasing first the inter-species interactions and then the intra-species one,  this limit is not reached and a ground-state wavefunction different from that of the three-boson system is realized. 

The paper is organized as follows. In Sec.~\ref{sec:Hamil}  we introduce the system Hamiltonian, discuss the spatial symmetries of the interaction potential and how the wavefunctions show a reduced symmetry due to the symmetrization condition over the two identical bosons. We introduce in this section analytical ansatzs valid for the possible limiting cases associated with the intra- and inter-species interactions. 
In Sec.~\ref{sec:numerical}  we obtain the whole energy spectra as a function of both the inter- and intra-species interactions. We discuss how the distinguishability of the third atom induces degeneracies associated to the absence of a symmetrization condition with respect to the other two atoms in the gas. In Sec.~\ref{sec:coherence} we discuss how coherences and  correlations are built in the system by the  inter- and intra-species interactions. Finally, we offer a summary and our conclusions in  Sec.~\ref{Sec:conc}.


\section{Hamiltonian of the system}
\label{sec:Hamil}

We consider a one-dimensional mixture of two identical bosons of one kind, $A$, with coordinates $x_1$ and $x_2$, and one atom
of kind $B$, with coordinate $y$. We assume the same mass $m$ and trapping oscillator frequency $\omega$ for the three atoms. We consider contact interactions modeled by a delta function of strength $g_{\mathrm{A}}$ between the A atoms and of strength $g_{\mathrm{AB}}$ between the A and B atoms. The constants $g_{\mathrm{A}}$ and $g_{\mathrm{AB}}$ are the intra- and inter-species coupling constants, respectively.  In this situation, the Hamiltonian reads  
\begin{align}
\label{eq:Hamiltonian}
&H =- \frac {1}{2} \frac {d^2}{dx_1^2} - \frac {1}{2} \frac {d^2}{dx_2^2} - \frac {1}{2} \frac {d^2}{dy^2} 
   + \frac {1}{2} x_1^2 +  \frac {1}{2} x_1^2 + \frac {1}{2} y^2\nonumber\\
&+ g_\mathrm{A} \delta (x_1 -x_2) + g_\mathrm{AB}  \delta (x_1-y) + g_\mathrm{AB}  \delta (x_2 -y) \,. 
\end{align}
To write Hamiltonian~(\ref{eq:Hamiltonian}) we  scaled all energies by $\hbar\omega$ and all distances by  the harmonic oscillator length $a_{\mathrm {ho}}=\sqrt{\hbar/m\omega}$.  Thus, all coupling constants are scaled by  $a_{\mathrm {ho}}\hbar\omega$.
Note that the eigenfunctions of Hamiltonian~(\ref{eq:Hamiltonian}) should  be symmetric with respect to 
the exchange of the $A$ bosons, with no symmetry restriction for the atom of type $B$.   
On top of this, Hamiltonian~(\ref{eq:Hamiltonian}) presents some {\it spatial} symmetries which can be elucidated by performing  the following Jacobi transformation:
\begin{align}
  \label{eq:transf}
& R=(x_1+x_2+y)/3, \nonumber\\
 &X=(x_1-x_2)/\sqrt{2}, \nonumber\\
 &Y=(x_1+x_2)/\sqrt{6}-\sqrt{2/3}y,
 \end{align} 
as introduced in~\cite{Harshman:12,Zinner:13}. In these variables, Hamiltonian~(\ref{eq:Hamiltonian}) becomes $H=H_{\mathrm{cm}}+H_{\mathrm{rel}}+V_{\mathrm{int}}$, with:
 \beal
  \label{eq:H_transf1}
 H_{\mathrm{cm}} &  = - \frac {1}{2} \frac {d^2}{dR^2}+ \frac {1}{2} R^2,\\
 H_{\mathrm{rel}}& =  - \frac {1}{2} \frac {\partial^2}{\partial X^2} - \frac {1}{2} \frac {\partial^2}{\partial Y^2} + \frac {1}{2}(X^2+Y^2),\nonumber\\
 V_{\mathrm{int}}& = g_{\mathrm{A}} \delta(X)+g_{\mathrm{AB}} \delta(-\frac{1}{2}X+\frac{\sqrt{3}}{2}Y) \nonumber \\
 &  + g_{\mathrm{AB}} \delta(-\frac{1}{2}X-\frac{\sqrt{3}}{2}Y).\nonumber
 \end{align}
  The first term of Hamiltonian~(\ref{eq:H_transf1}),  $H_{\mathrm{cm}}$, describes the motion of the center of mass, with coordinate $R$, which corresponds to that of a single particle of mass $M=3m$. The second and third terms, $H_{\mathrm{rel}}$ and  $V_{\mathrm{int}}$,  are associated with the relative motion of the three atoms, which occurs in the plane defined by $X$ and $Y$. 
  Let  $|\nu,\mu,\eta\rangle $ be an eigenfunction of the single-particle part  $H_\mathrm{sp}=H_{\mathrm{cm}}+H_{\mathrm{rel}}$. The center-of-mass motion separates from the relative motion, so that $|\nu,\mu,\eta\rangle= |\eta\rangle\otimes|\nu,\mu\rangle$, with $|\eta\rangle$ being the one-dimensional harmonic oscillator eigenstates, expressed in coordinate space as 
  
  \begin{equation}
 \psi_\eta(R)=\pi^{-1/4}(2^\eta\eta!)^{-1/2}H_\eta(R)\exp(-R^2/2),\label{eq:hermite}  
  \end{equation}
  $H_\eta(R)$ being the Hermite polynomials.

\begin{figure*}
\includegraphics[width=1.8\columnwidth]{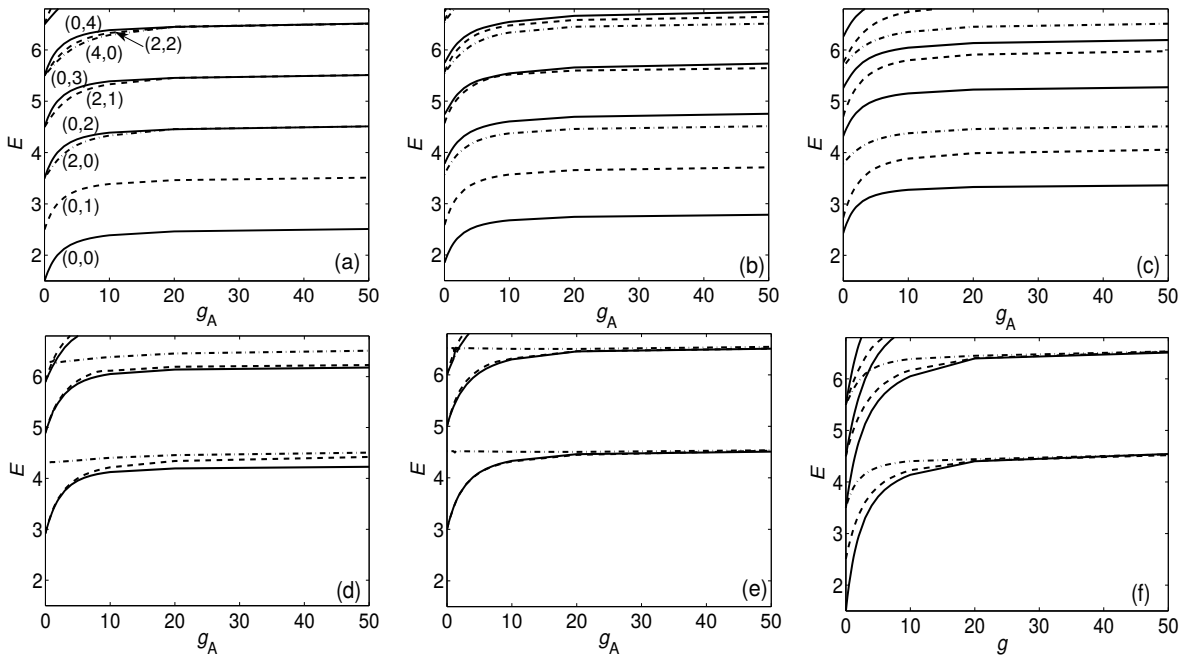}
\caption{(a) to (e) energy eigenspectrum as a function of $ g_{\mathrm{A}}$ for  $g_{\mathrm{AB}} =0,0.5,2,10,50$, respectively. (f)  energy eigenspectrum as a function of $g_{\mathrm{AB}} =g_{\mathrm{A}}=g$. Different line styles are used to help identify the different excited states. In panel a), the values of  the quantum numbers $(n_X,n_Y)$, which fully characterize the state at $ g_{\mathrm{A}}=0$,  are indicated. Note that they also correspond to the number of nodes in each direction $X$ and $Y$ for non-zero but small values of  $g_{\mathrm{A}}$ (see Fig.~\ref{fig6}). We use harmonic oscillator units for the energies and distances. \label{fig1}}
\end{figure*}

 \subsection{Symmetries of the Hamiltonian}
 \label{sec:symmetry}
 
 The kinetic and external trapping potential terms in Hamiltonian~(\ref{eq:H_transf1}) show continuous cylindrical symmetry. As we show below, the interaction term, $V_{\mathrm{int}}$, has certain discrete rotational symmetry in the $X$-$Y$ plane. In addition to these symmetries of the Hamiltonian, the symmetrization of the identical A bosons imposes an additional constrain on the wavefunctions, as they have to be invariant under the interchange of the two A atoms~\cite{Harshman:12}. Let us see how this discrete symmetry of $V_{\mathrm{int}}$ together with the symmetrization condition over the identical bosons permit us to grasp some properties of the wavefunctions of the system. 
 
 {\it Non-interacting case} -- In the absence of interactions, the relative coordinate part of   Hamiltonian~(\ref{eq:H_transf1}), $H_{\mathrm{rel}}$, shows continuous cylindrical symmetry. 
 Indeed, $H_{\mathrm{rel}}$ is separable in the $X$ and $Y$ directions and, therefore, the solutions in the relative coordinates  can be written as products of the harmonic oscillator eigenstates  $\psi_{n_X}(X)$ and $\psi_{n_Y}(Y)$, which have the form of Eq.~(\ref{eq:hermite}). 
Eigenfunctions $\psi_{n_X}(X) $ and $\psi_{n_Y}(Y)$  are characterized by two quantum numbers, $n_X$ and $n_Y$. While $n_Y$ can take any integer value, $n_X$ is restricted to even numbers. This is a consequence of  the symmetrization of the identical A bosons.  The interchange of the two A atoms is isomorphic to  the spatial reflection $X\to-X$, that is  a reflection with respect to the $Y$ axis. Since these two atoms are identical, the wavefunctions have to remain unchanged under their interchange, or, what is the same the transformation  $X\to-X$. In particular that means that it cannot change sign, and therefore, $n_X$ has to be even. 

 {\it Interacting case} -- To find the spatial symmetry of  $V_{\mathrm{int}}$ we perform a new transformation to cylindrical coordinates in the $X$-$Y$ plane
\begin{align}
  \label{eq:transf2}
 &\rho=\sqrt{X^2+Y^2}, \nonumber\\
 &\tan \phi=Y/X,
 \end{align} 
which leads to
 \begin{align}
  V_{\mathrm{int}} &= g_{\mathrm{A}} \delta(\rho\cos\phi) + g_{\mathrm{AB}} \delta(-\frac{1}{2}\rho\cos\phi+\frac{\sqrt{3}}{2}\rho\sin\phi)\nonumber\\ 
 &+ g_{\mathrm{AB}} \delta(-\frac{1}{2}\rho\cos\phi-\frac{\sqrt{3}}{2}\rho\sin\phi).
 \end{align} 
 This interaction potential can be simplified to $V_{\mathrm{int}}^\mathrm{A}+V_{\mathrm{int}}^{\mathrm{AB},+} +V_{\mathrm{int}}^{\mathrm{AB},-}$, with
 \beq
  \label{eq:int_pot_A}
 V_{\mathrm{int}}^A =(g_{\mathrm{A}}/|\rho|)\left[\delta(\phi-\pi/2)+\delta(\phi-3\pi/2)\right],
 \eeq
 and
 \beq
  \label{eq:int_pot_AB}
 V_{\mathrm{int}}^{AB,\pm} =(g_{\mathrm{AB}}/|\rho|)\left[\delta(\phi\mp\pi/6)+\delta(\phi\pm5\pi/6) \right],
 \eeq
where we assume $|\rho|\neq0$ (see footnote 2 in Ref.~\cite{Harshman:12}). For $g_{\mathrm{AB}}\ne g_{\mathrm{A}}$ the interaction potential shows $\mathcal{C}_{2v}$ symmetry. For $g_{\mathrm{AB}}=g_{\mathrm{A}}=g$
it shows $\mathcal{C}_{6v}$ symmetry, as it reads
\beq
  \label{eq:int_pot_total}
 V_{\mathrm{int}} = (g/|\rho|)\sum_{i=1}^6\delta(\phi-\frac{2i-1}{6}\pi).
\eeq
The discrete rotational symmetry of $V_{\mathrm{int}}$ imposes that the wavefunctions have to belong  to the irreducible representations either of the discrete groups conventionally termed  as $\mathcal{C}_{6v}$  (if $g_{\mathrm{AB}}= g_{\mathrm{A}}$) or  to $\mathcal{C}_{2v}$ (if $g_{\mathrm{AB}}\ne g_{\mathrm{A}}$)~\cite{Hamermesh}. In discrete rotational group theory, the different discrete transformations associated with certain discrete groups are called group elements. For example, the reflection with respect to certain axis  is a group element conventionally denoted as  $\sigma_\nu$~\cite{Hamermesh}. Normally, there  are, at most, two non-equivalent axis, and the second one is denoted as $\sigma_d$. Then, there can be a number of equivalent axis, which are denoted as $\sigma_{\nu'}, \sigma_{\nu''},\dots$ or $\sigma_{d'},\sigma_{d''},\dots$ The discrete rotations of an angle $\pi/k$ are denoted as $\mathcal{C}_k$, and together with the reflections complete the possible elements associated to a particular discrete rotational group. 

 The number of irreducible representations of a discrete group of finite order, like  $\mathcal{C}_{6v}$ or   $\mathcal{C}_{2v}$, is finite. The irreducible representation to which  a particular wavefunction belongs  determines how this function is transformed under the action of the elements of the group. 

In the system of two A identical atoms and a third B atom, some irreducible transformations of $\mathcal{C}_{6v}$ or   $\mathcal{C}_{2v}$   are forbidden as a consequence of the symmetrization of the identical A bosons.
Then, the wavefunctions can only belong to irreducible representations of $\mathcal{C}_{6v}$ or   $\mathcal{C}_{2v}$  which remain unchanged under the transformation $X\to-X$, that is, which are transformed in a particular way under the action of the reflection $\sigma_\nu$. For $\mathcal{C}_{6v}$  ($g_{\mathrm{AB}}= g_{\mathrm{A}}$) the irreducible representations of   $\mathcal{C}_{6v}$ that do not change sign under this reflection are  $\mathcal{A}_1$ or $\mathcal{B}_2$ (we use the conventional terminology to name the irreducible representations, see e.g. ~\cite{Hamermesh}). Then,  the actual wavefunctions have to belong to any of these two representations. 

The fact that some wavefunctions which, in principle, could exist according to the symmetry of the interaction potential are forbidden due to the the symmetrization condition occurs also in  the non-interacting case. In that case, the symmetrization condition is responsible for   the forbidding of the wavefunctions with  odd values of $n_X$. 

There is no restriction with respect to the other two $\sigma_{\nu',\nu''}$ axis which are equivalent  to the $Y$ axis. These are the lines passing through the origin with an angle $\pm\pi/6$.  Also, there is no restriction regarding the reflections with respect to the $\sigma_d$ axis, which are the $X$ axis and  the lines passing through the origin with an angle $\pm\pi/3$. Discrete rotations $\mathcal{C}_2$  of an angle $\pi$ or  reflections $\sigma_d$  have an associated character of 1 or -1 for  the irreducible representations  $\mathcal{A}_1$ or $\mathcal{B}_2$ of both  $\mathcal{C}_{6v}$ and $\mathcal{C}_{2v}$. 

All elements of the group $ \mathcal{C}_{6v}$ are isomorphic to a permutation of the three atoms (see~\cite{Harshman:12}), and therefore to a spatial transformation. For example, 
 the permutation of the distinguishable atom with coordinate $y$ with one A atom, say $x_1$, is isomorphic to the {\it spatial } transformation
 \begin{align}
  \label{eq:transf2}
 &X'=X/2+\sqrt{3}Y/2, \nonumber\\
 &Y'=\sqrt{3}X/2-Y/2,
 \end{align} 
which is a reflection with respect to the  axis with  angle $-\pi/6$.  The permutation  of the B atom with the A atom at $x_1$, together with the permutation of this atom with the one at $x_2$ is isomorphic to the transformation
\begin{align}
  \label{eq:transf2}
 &X'=-X/2-\sqrt{3}Y/2, \nonumber\\
 &Y'=\sqrt{3}X/2-Y/2.
 \end{align} 
 which is a $ \mathcal{C}_{3}$ rotation. 
 These are spatial transformations which leave Hamiltonian~(\ref{eq:H_transf1}) unchanged. In short, under all possible transformations associated to the $ \mathcal{C}_{6v}$ group,  wavefunctions have to transform in a way compatible with the   symmetrization of the identical A bosons.  This is obeyed by the wavefunctions belonging to the  $\mathcal{A}_1$ or $\mathcal{B}_2$ irreducible representations. We stress here  that the same two representations are the permitted ones for the $ \mathcal{C}_{2v}$ group, when $g_{\mathrm{AB}}\ne g_{\mathrm{A}}$. 
 
 This is a non-trivial way of reducing the  $ \mathcal{C}_{6v}$ (or $ \mathcal{C}_{2v}$ if $g_{\mathrm{AB}}\neq g_{\mathrm{A}}$) symmetry. Particularly, the absence of a symmetrization condition between the A and B atoms implies that there is no condition over the  sign of the wavefunction when crossing the lines  $X=\pm\sqrt{3}Y$, that is, the axis with slope $\pm\pi/6$. These axis are the locus of the points at which the B atom locates at the same position than one of the A atoms. 
This 
is of particular relevance in the limiting cases in which the coupling constants either tend to infinity or vanish. 

 The  $\mathcal{A}_1$ or $\mathcal{B}_2$  irreducible representations are of one dimensional type.  The group $ \mathcal{C}_{6v}$ has some representations which are two-dimensional, but all of them are forbidden due to the symmetrization condition. Then, any degeneracy that occurs when considering only the relative motion in the $X$-$Y$ plane is not associated to the discrete symmetry. As we will show, some degeneracies occur in the spectra when any coupling constant tends to zero or to infinity, that is,  in the following extreme limits: 
 \begin{enumerate}[label=(\roman*)]
  \item when all coupling constants are zero
  \item when $g_{\mathrm{A}}\to\infty$,  $g_{\mathrm{AB}}=0$,
  \item when $g_{\mathrm{AB}}\to\infty$,  $g_{\mathrm{A}}=0$, and
  \item when $g_{\mathrm{A}}\to\infty$,  $g_{\mathrm{AB}}\to\infty$.
 \end{enumerate}
When all coupling constants vanish [(case (i)],  the ground state is non-degenerate, and degeneracies occur only in the excited states. We will show that in the rest of cases degeneracies occur also in the ground state. 
 On the other hand, for (i), the wavefunction is trivially given by products of the eigenfunctions of the one-dimensional harmonic oscillator, as discussed above.
For  $g_{\mathrm{A}}\to\infty $, $g_{\mathrm{AB}}=0 $, or for $g_{\mathrm{AB}}\to\infty $, $g_{\mathrm{A}}=0 $, exact analytical forms of the wavefunctions, can be obtained~\cite{Harshman:12,Zinner:13}. The first limit is the fermionization of two bosons~\cite{Girardeau:01}. The second one is a composite fermionized gas~\cite{Zollner:08a}, the smallest mixture that one can consider for imbalanced number of atoms in the two species. Finally, if both $g_{\mathrm{A}}$ and $g_{\mathrm{AB}}$ $ \to\infty  $,  analytical forms of the wavefunctions can be proposed as well~\cite{Girardeau:07,Harshman:12,Zinner:13}. We will  use the fact that the wavefunctions should  be symmetric with respect to  the exchange of the $A$ bosons, with no symmetry restriction for the atom of type $B$, to propose ansatzs  in all these limits.  These ansatzs will help to understand the origin of the degeneracies in each limit and will also serve to add physical insight to the solutions obtained in Refs.~\cite{Harshman:12,Zinner:13}. 
Eventually, they can also be used as trial functions for Quantum  Monte Carlo calculations.

In the following, we use  theoretical  arguments and numerical calculations  to show that the degeneracies that occur in these limits are related to the absence of  a symmetrization condition, i.e. when the particles are distinguishable. 

\begin{figure*}
\includegraphics[width=1.85\columnwidth]{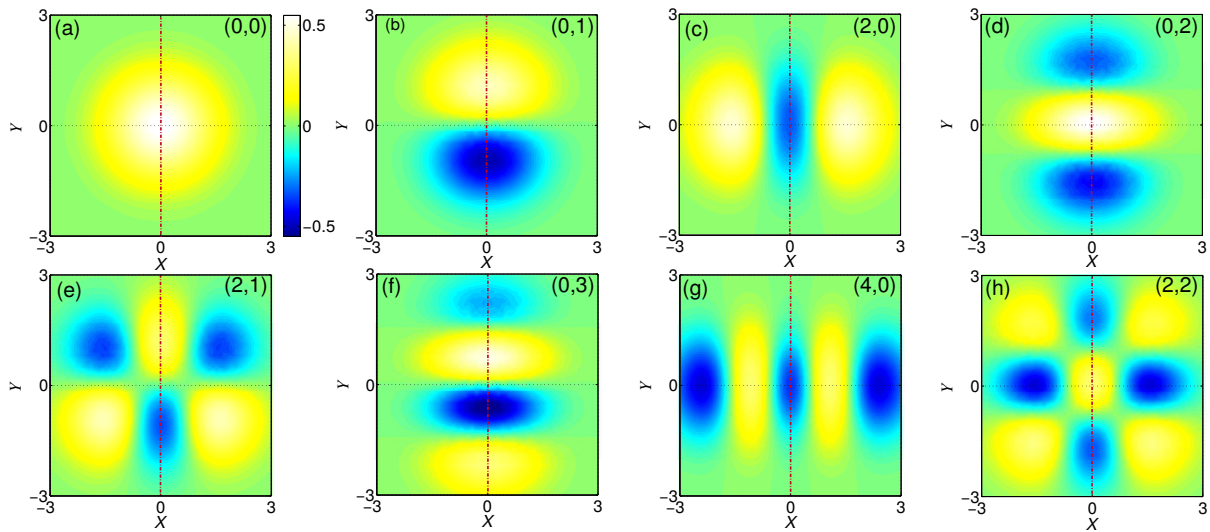}
\caption{(Color online) {\it  Wavefunctions of the ground and excited states in the $X$-$Y$ plane, when  $g_{\mathrm{A}} =0.1$ and $g_{\mathrm{AB}} =0$}. (a) and (b)  represent the ground and first excited energy wavefunctions. (c)  represents the excited state with $(n_X,n_Y)=(2,0)$ which is quasi-degenerate with (d), for which $(n_X,n_Y)=(0,2)$.  (e) is the excited state with   $(n_X,n_Y)=(2,1)$, quasi-degenerate with  (f) which has $(n_X,n_Y)=(0,3)$. (g) and (h) are the quasi-degenerate excited states with  $(n_X,n_Y)=(4,0)$   and  $(n_X,n_Y)=(2,2)$, respectively, which are also quasi-degenerate with $(n_X,n_Y)=(0,4)$ (not shown).  The dash-dotted red line highlights the axis along which the  interactions among the A atoms occur.  \label{fig6}}
\end{figure*}

\subsection{Analytical ansatzs}
\label{sec:degeneracies}

In this section, we do not restrict our analysis to the relative motion in the  $X$-$Y$ plane, but consider the total Hamiltonian~(\ref{eq:Hamiltonian}) in terms of the positions of each atom $(x_1,x_2,y)$.  
In the non-interacting case, the ground state  is non-degenerate, and its wavefunction is symmetric under the exchange of all atoms, real, positive and without zeros. It is given by
\beq
\Psi_{\rm g.s.}(x_1,x_2,y)\!  =\!  \psi_0(x_1) \psi_0(x_2) \psi_0(y) \,.
\eeq
The energy of the ground state is 
$E_{\rm g.s.}= 3/2$.

In the second limit, that is, when $g_{\mathrm{A}}\to\infty$,  $g_{\mathrm{AB}}=0$, the two A atoms form a TG gas, while the single atom in B does not interact with  this gas. The ground state is known  to be
\begin{align}
\Psi_{\rm g.s.}^{1,\rm bos}(x_1,x_2,y)  
& \propto\exp{[ - \frac {1}{2} (x_1^2+x_2^2+y^2)]}\times |x_1-x_2|.
\label{eq:analyticalwf_2}
\end{align}

The third limit, when $g_{\mathrm{AB}}\to\infty $, $g_{\mathrm{A}}=0 $, is the composite fermionization limit. The following wavefunction has been proposed to be a good ansatz for the wavefunction~\cite{Zollner:08a,Garcia-March:13}
\begin{align}
\label{eq:analyticalwf_3}
\Psi_{\rm g.s.}^{2,\rm bos}(x_1,x_2,y)  
&\propto \exp{[ - \frac {1}{2} (x_1^2+x_2^2+y^2)]}  \\
&\times |x_1-y|\; |x_2-y|.\nonumber
\end{align}
This wavefunction is zero whenever one atom of A and the B atom  are in the same position. In the relative motion plane, this occurs along the lines $X=\pm\sqrt{3} Y$.  This function is real and positive, that is, it has zeros but not changes sign. As there is no symmetrization condition  between A and B atoms the following wavefunction is equally a possible ansatz
\begin{align}
\Psi_{\rm g.s.}^{1,\rm bos}(x_1,x_2,y)  
& \propto\exp{[ - \frac {1}{2} (x_1^2+x_2^2+y^2)]}\nonumber\\
&\times  (x_1-y)(x_2-y).
\label{eq:analyticalwf_3b}
\end{align}
Now, this form of the wavefunction permits changes of sign. Wavefunctions~(\ref{eq:analyticalwf_3}) and ~(\ref{eq:analyticalwf_3b}) are not orthogonal but are degenerate in energy. Another possibility is the following wavefunction
\begin{align}  
\Psi_{\rm g.s.}^{3,\rm bos}(x_1,x_2,y)  
&\propto \exp{[ - \frac {1}{2} (x_1^2+x_2^2+y^2)]} \nonumber\\
&\hspace{-1cm}\times \left[ (x_1-y)\; |x_2-y| + |x_1-y|(x_2-y)\right]. 
\label{eq:analyticalwf_3c}
\end{align}
 This wavefunction is orthogonal to the previous two  and it is degenerate in energy with them. In  the relative motion plane, it  has zeros and changes of sign along the lines $X=\pm\sqrt{3} Y$. Up to know, we cannot assess wether the ground state in this limit is non-degenerate,  two- or three-fold degenerate. We will show in the next section that the ground state is indeed two-fold degenerate when $g_{\mathrm{AB}}\to\infty$ and  $g_{\mathrm{A}}=0$.   

Finally, in the limit where both coupling constants are infinite, the wavefunction can be  obtained by  constructing 
a  Slater determinant (to avoid two atoms to be in the same point)  with the harmonic oscillator single-particle wavefunctions, expand the Vandermonde determinant,  
and  add absolute values to the products of the binomials in the difference of coordinates 
to obtain the desired symmetries. In all cases  one should assure that 
the wavefunction is zero whenever two atoms are in the same position, in such a way that the interaction 
energy is zero.  For instance, one of the possible   wavefunctions  is
\begin{align}
\label{eq:analyticalwf_4}
\Psi_{\rm g.s.}^{1,\rm bos}(x_1,x_2,y)  
& \propto\exp{[ - \frac {1}{2} (x_1^2+x_2^2+y^2)]}\nonumber\\
&\times |x_1-x_2| (x_1-y)(x_2-y),
\end{align}
which is symmetric under the exchange of A atoms and has energy 
 $E= 1/2+3/2+5/2$.  This function is real, has zeros and
 changes sign.  Indeed, it has zeros along the lines in which the two A atoms are in the same position (the $Y$ axis in the relative motion plane) and when  any A atom is in the same position than the B atom (the lines $X=\pm\sqrt{3} Y$ in the relative motion plane). It has  changes of sign when crossing the lines $X=\pm\sqrt{3} Y$ of the  relative motion plane. 

\begin{figure}
\includegraphics[width=0.95\columnwidth]{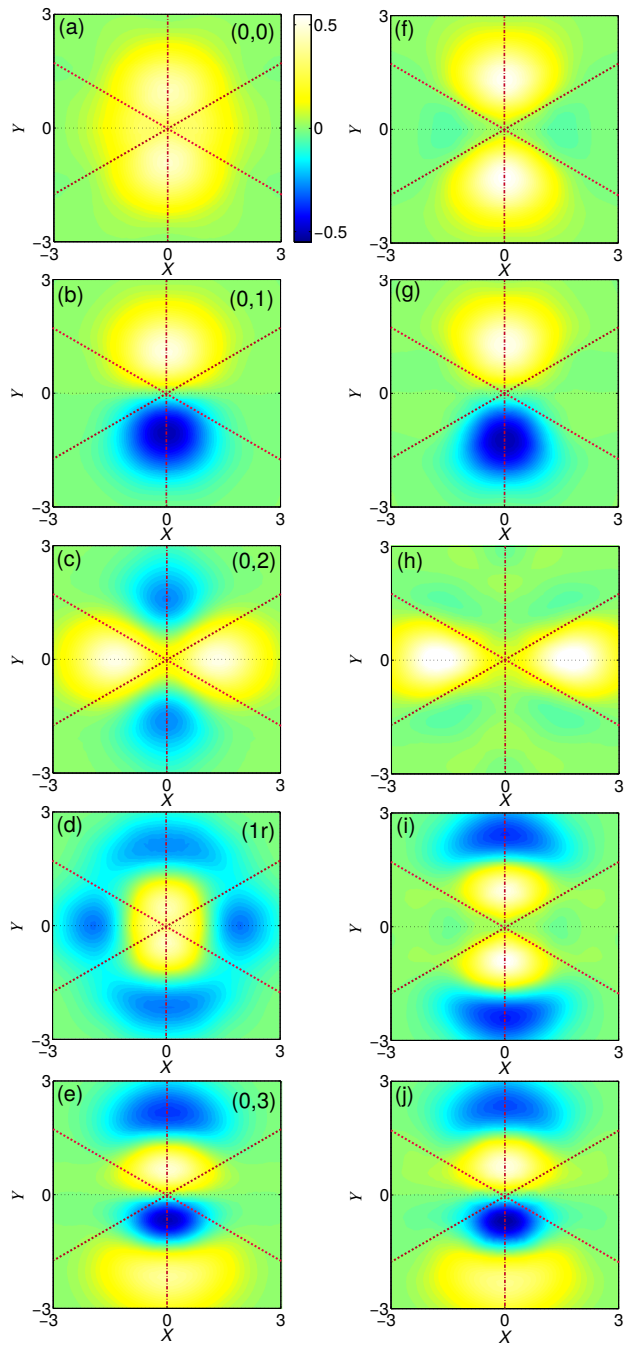}
\caption{(Color online) {\it  Wavefunctions of the ground and excited states in the $X$-$Y$ plane, when  $g_{\mathrm{A}} =0$ and $g_{\mathrm{AB}} =2$ and 10}. (a) to (e) represent the lowest five states for  $g_{\mathrm{AB}} =2$, and (f) to (j) for  $g_{\mathrm{AB}} =10$. The ground and first excited states are quasi-degenerate for $g_{\mathrm{AB}} =10$. Also the third and fourth states are quasi-degenerate. Conversely, the second excited state is non-degenerate. Dashed (dash-dotted) red  lines highlight the axis along the which the AB (A) interactions occur.  In panels (a) to (e), we indicate the number of nodes in the $X$ ($N_X$)  and $Y$ ($N_Y$) directions  as $(N_X,N_Y)$. Instead, when $N$ radial nodes occur, we indicate them as $(Nr)$. \label{fig2}}
\end{figure}

Another possible wavefunction, degenerate in energy with the previous one,  is 
the completely symmetrized version of the  Slater determinant
\begin{align}
\label{eq:analyticalwf_4b}
\Psi_{\rm g.s.}^{2,\rm bos}(x_1,x_2,y)  
&\propto \exp{[ - \frac {1}{2} (x_1^2+x_2^2+y^2)]}  \\
&\times |x_1-x_2| \;|x_1-y|\; |x_2-y|.\nonumber
\end{align}
 This wavefunction  has again zeros along the  $Y$ axis and the lines $X=\pm\sqrt{3}Y$, but the sign is not changed when crossing these lines. 
Finally, a third possible wavefunction is
\begin{align}  
\Psi_{\rm g.s.}^{3,\rm bos}(x_1,x_2,y)  &\propto \exp{[ - \frac {1}{2} (x_1^2+x_2^2+y^2)]}|x_1-x_2|\nonumber \\
&\hspace{-1cm}\times \left[ (x_1-y)\; |x_2-y| + |x_1-y|(x_2-y)\right]. 
\label{eq:analyticalwf_4c}
\end{align}
This wavefunction also  fulfills  the requirement that the wavefunction should be  zero whenever
 two atoms are in the same position and has the same energy than the previous ones. 
Again, we cannot, at this point, elucidate which is the degeneracy of the ground state. We will show in the next section that the ground state is three-fold degenerate when $g_{\mathrm{A}}$ and $g_{\mathrm{AB}}\to\infty$. 
 
 Note that,  
wavefunction~(\ref{eq:analyticalwf_4c}) is orthogonal to wavefunctions~(\ref{eq:analyticalwf_4}) and~(\ref{eq:analyticalwf_4b}), but wavefunctions~(\ref{eq:analyticalwf_4}) and~(\ref{eq:analyticalwf_4b}) are not orthogonal among themselves. In the following section we evaluate numerically the validity of the ansatzs proposed in this section, and interpret them in view of the symmetry analysis of section~\ref{sec:symmetry}.

\begin{figure}
\includegraphics[width=0.95\columnwidth]{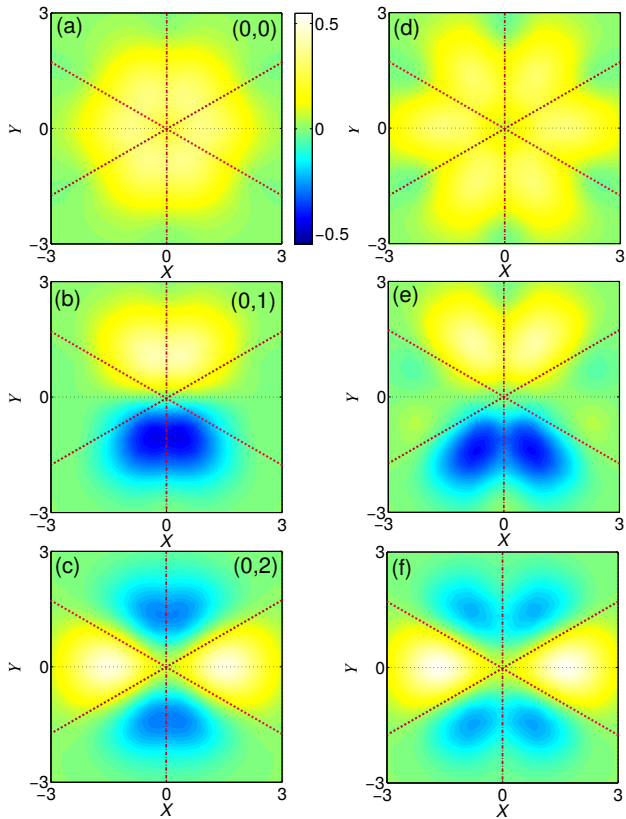}
\caption{(Color online) {\it  Wavefunctions of the ground and excited states in the $X$-$Y$ plane, when  $g_{\mathrm{A}} =g_{\mathrm{AB}}$}. (a) to (c) represent the three lowest energy  wavefunctions when  $g_{\mathrm{A}} =g_{\mathrm{AB}}=2$ and while (d) to (f) the ones when $g_{\mathrm{A}} =g_{\mathrm{AB}}=10$.  Dashed (dash-dotted) red  lines highlight the axis along the which the AB (A) interactions occur. In panels (a) to (c), we indicate the number of nodes in the $X$ ($N_X$)  and $Y$ ($N_Y$) directions  as $(N_X,N_Y)$.   \label{fig3}}
\end{figure}

\section{Distinguishability and degeneracies in the energy spectra} 
\label{sec:numerical}


To numerically evaluate the energy spectra of Hamiltonian~(\ref{eq:Hamiltonian})  we employ the exact diagonalization algorithm described in~\cite{Garcia-march:12} and outlined in appendix~\ref{sec:numericalmethod}.

\subsection{Variable intra-species interactions with vanishing inter-species interactions}

We show in Fig.~\ref{fig1} the energy spectra as a function of $g_{\mathrm{A}}$, for different values of   $g_{\mathrm{AB}}$.  We plot  the total energies, that is including that of the center of mass.  We nevertheless only plot the energies of excited states which correspond to excitations in the relative motion, and not in the center of mass motion.  
 In Fig.~\ref{fig1}(a) we show the energy spectra when  $g_{\mathrm{AB}} =0$. This is the superposition of the energy spectra of the motion of two atoms, analytically calculated in~\cite{Busch:98}, and that of the single distinguishable atom.  There are many degeneracies due to the presence of this extra atom. As discussed in Sec.~\ref{sec:symmetry}, for $g_{\mathrm{A}}=g_{\mathrm{AB}}=0$, the solutions can be written as products of the one-dimensional eigenfunctions of the harmonic oscillator in the relative motion $X$-$Y$ plane, written in terms of Hermite polynomials as in Eq.~(\ref{eq:hermite}), and characterized by two quantum numbers, $(n_X,n_Y)$, $n_X$ only taking even values. The total energy of each eigenfunction is $E=3/2+n_X+n_Y$. The excited states with equal value of the sum $n_X+n_Y$ are 
degenerate when $g_{\mathrm{A}}=0$. The 
energies of these states are the ones shown in Fig.~\ref{fig1} (a). For example,  the first excited state is non-degenerate and has $(n_X,n_Y)=(0,1)$, while the second and third excited states are degenerate at $g_{\mathrm{A}}=0$ and have $(n_X,n_Y)=(0,2)$ and $(n_X,n_Y)=(2,0)$. The fourth and fifth states are again degenerate and have $(n_X,n_Y)=(2,1)$ and $(n_X,n_Y)=(0,3)$, while the next three degenerate states have $(n_X,n_Y)=(4,0)$, $(n_X,n_Y)=(0,4)$, and $(n_X,n_Y)=(2,2)$.  The number of degenerate states observed in Fig.~\ref{fig1}(a) for  $g_{\mathrm{A}}=0$ correspond to the possible values of $(n_X,n_Y)$ fulfilling   $n_X+n_Y=0,1,2,3$ or 4.

For $g_{\mathrm{A}}=0$ and for those states which are degenerate, any  linear combination of the eigenfunctions written in the form of   Eq.~(\ref{eq:hermite}) is also a possible solution. On the contrary, 
for small, finite $g_{\mathrm{A}}$, this degeneracy is broken. Since the Hamiltonian contains only the interacting part in the $X$ direction, given by Eq.~(\ref{eq:int_pot_A}) which  reads $g_{\mathrm{A}} \delta(X)$, the non-degenerate eigenstates for small $g_{\mathrm{A}}$ are still well described by the quantum numbers $(n_X,n_Y)$. 
The first eight eigenfunctions   for small $g_{\mathrm{A}}$ are shown in Fig.~\ref{fig6}. To obtain these profiles, we numerically calculate the wavefunction in second quantization by direct diagonalization, and transform them back to  first quantization, as described in Appendix~\ref{sec:numericalmethod}. All these wavefunctions  are very well approximated by products of the eigenfunctions given by   Eq.~(\ref{eq:hermite}), with $(n_X,n_Y)$ up to $n_X+n_Y=4$.  

For $g_{\mathrm{A}}\ne0$, the interaction potential raised at the $Y$ axis, which is described by Eq.~(\ref{eq:int_pot_A}), shows $\mathcal{C}_{2v}$ symmetry. This symmetry  is reduced by the symmetrization condition, so that all wavefunctions in Fig.~\ref{fig6} belong to the $\mathcal{A}_1$ or $\mathcal{B}_2$ representations (and then show parity 1 or -1). Moreover, we note that for small but finite values of $g_{\mathrm{A}}$, the energy of the state $(n_X,n_Y)=(0,2)$ is larger than that of state $(n_X,n_Y)=(2,0)$ [see Fig.~\ref{fig1}(a)]. Similarly, the energy of state  $(n_X,n_Y)=(0,4)$ is larger than the one of state  $(n_X,n_Y)=(2,2)$, which is larger than that of state $(n_X,n_Y)=(4,0)$. In general, among those states which are degenerate for  $g_{\mathrm{A}}=0$, those with larger $n_Y$ have larger energy for finite  $g_{\mathrm{A}}$.   This is  due to the fact that  the states with larger $n_Y$ show larger non-zero values of the wavefunction around the $Y$ axis (see  Fig.~\ref{fig6}), which is where 
interaction potential occurs.  

As $g_{\mathrm{A}}$ is the strength of a delta potential along the $X=0$ line, when  $g_{\mathrm{A}}\ne0$ these solutions are deformed along this line until a zero is reached 
along the $Y$ axis for $g_{\mathrm{A}}\to\infty$.   The degeneracy is recovered for large $g_{\mathrm{A}}$, when a zero occurs along this line. In the following, we use the spectra shown  in Fig.~\ref{fig1}(a) as a benchmark to understand the system when $g_{\mathrm{AB}}\ne0$.

\begin{figure}
\includegraphics[width=0.68\columnwidth]{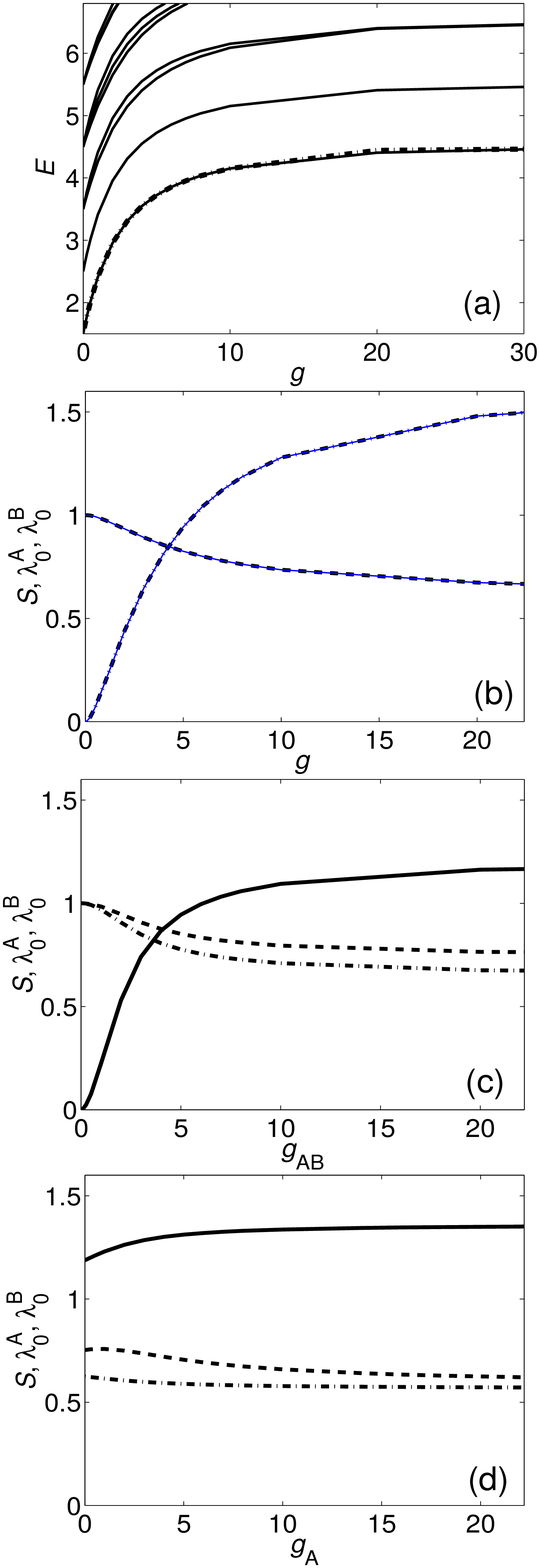}
\caption{(Color online)  (a) Energy spectra for three indistinguishable atoms  (solid black lines). The dash-dotted blue line  is the energy of the ground state for the system of two A atoms and a third distinguishable B atom, when   $g_{\mathrm{A}}=g_{\mathrm{AB}}=g$.  This line overlaps with the one corresponding to the ground state of the system of three indistinguishable atoms. (b) Largest occupation of a natural orbit for three indistinguishable atoms (dashed black line) and von Neumann Entropy  (dash-dotted black line). Thin blue lines overlapping with these two curves are the same quantities calculated for the system of two A atoms and a third B atom, when $g_{\mathrm{A}}=g_{\mathrm{AB}}=g$. In (c)  we plot the largest  occupation of a natural orbit for A and B (dash-dotted and dashed lines, respectively), when $g_{\mathrm{AB}}$ is increased, keeping $g_{\mathrm{A}}=0$. We also plot in (c) the von Neumann entropy (solid line). The same quantities are plotted in (d), when  $g_{\mathrm{AB}}=50$ and $g_{\mathrm{A}}$ is increased. 
 \label{fig4}}
\end{figure}

 \subsection{Variable inter-species interactions with vanishing intra-species interactions}

The degeneracies occurring for $g_{\mathrm{A}}=g_{\mathrm{AB}}=0$ are also lifted for finite $g_{\mathrm{AB}}$, as shown in Fig.~\ref{fig1} (b) and (c), where we plot  the energy for the ground states and the aforementioned  excitations of the relative motion for $g_{\mathrm{AB}}=0.5$ and 2, respectively, as a function of  $g_{\mathrm{A}}$.  The AB interactions occur along the $X=\pm\sqrt{3}Y$ lines, with strength $g_{\mathrm{AB}}$ [see Eq.~(\ref{eq:int_pot_AB})]. Then, the solutions cannot be expressed in general as products of the functions~(\ref{eq:hermite}). We represent in the first column of  Fig.~\ref{fig2} the wavefunctions up to the fourth excited state for $g_{\mathrm{AB}}=2$ and  $g_{\mathrm{A}}=0$. The interaction potential shows $\mathcal{C}_{2v}$ symmetry but, again as a consequence of the symmetrization condition, 
 all wavefunctions in Fig.~\ref{fig2} belong to the $\mathcal{A}_1$ or $\mathcal{B}_2$ representations.  We show in the first column of  Fig.~\ref{fig7} of  Appendix~\ref{sec:appendixB}, the wavefunctions for the 
 fifth to seventh excitations, which also belong either to  $\mathcal{A}_1$ or to the $\mathcal{B}_2$ representation. In the first column of Figs.~\ref{fig2} and~\ref{fig7}  we indicate the number of nodes in the $X$ and $Y$ direction. While in some cases they resemble the quantum numbers $(n_X,n_Y)$, in general these quantum numbers are not good quantum numbers for $g_{\mathrm{AB}}\ne0$. Indeed,  we observe that the  states that show  $n_X\ne0$ and $n_Y=0$  when $g_{\mathrm{AB}}=0$ and finite $g_{\mathrm{A}}$ (see panels (c) and (g) in Fig.~\ref{fig6}), present radial nodes for  $g_{\mathrm{A}}=0$ and finite $g_{\mathrm{AB}}$ (see panel (d) in Fig.~\ref{fig2} and panel (c) in Fig.~\ref{fig7}). 
  
For  $g_{\mathrm{AB}}\to\infty$, the wavefunctions develop a zero along the  $X=\pm\sqrt{3}Y$ lines.  In the second column of Fig.~\ref{fig2} we report  the wavefunctions for $g_{\mathrm{AB}}=10$ and  $g_{\mathrm{A}}=0$. They are in agreement with the analytical solutions found for this limit in Ref.~\cite{Zinner:13}. We plot the  energies for $g_{\mathrm{AB}}=10$ in Fig.~\ref{fig1} (d). For $g_{\mathrm{A}}=0$ and  large  $g_{\mathrm{AB}}$, the first two states are quasi-degenerate. 
Each of these states belong either to  $\mathcal{A}_1$ or $\mathcal{B}_2$ representations, thus showing parity 1 or -1 [see Figs.~\ref{fig2} (f) and (g)].   The third and fourth states are also quasi-degenerate. They  are radial excitations along the $Y$ direction of the quasi-degenerate ground and first excited state [compare Figs.~\ref{fig2} (f) and (g) with Figs.~\ref{fig2} (i) and (j)]. Similarly occurs with the fifth and sixth excited states, shown in panels (d) and (e) of  Fig.~\ref{fig7}, which are excitations of  the states represented in Figs.~\ref{fig2} (f) and (g) showing an even number of nodes in the horizontal direction. The second excited state, represented in  panel (h) of Fig.~\ref{fig2},  is non-degenerate. Similarly occurs for the seventh excited state   (panel (f) of Fig.~\ref{fig7}), which is a radial excitation along the $X$ direction of the second excited state [compare Fig.~\ref{fig2} (h) and  Fig.~\ref{fig7}) (f)]. Both states show even parity. 

The solutions obtained in Ref.~\cite{Zinner:13} for $g_{\mathrm{AB}}\to\infty$ and $g_{\mathrm{A}}=0$  include states with either both   integer or fractional relative energy. These energies are reproduced for very strong repulsion (see Fig.~\ref{fig1} (e), where we plot the total energies of the ground and excited states for $g_{\mathrm{AB}}=50$).  Thus, the second and seventh states are the integer energy states, with relative energy $E=4 $ and $6 $, respectively, while the rest are fractional energy states. 

Finally, we find numerically that ground and first excited states have an overlap larger than $0.9$ with the ansatzs~(\ref{eq:analyticalwf_3}) and~(\ref{eq:analyticalwf_3c}), respectively. The overlap is obatined through an integral in the space defined by the spatial coordinates of the three atoms, $x_1$, $x_2$, and $y$. As it requires a grid in this space and numerical integration, this overlap is not exact. On the other hand, when transforming the ansatzs~(\ref{eq:analyticalwf_3}) and~(\ref{eq:analyticalwf_3c}) to the relative coordinates and plotting them in the $X$-$Y$ plane, we find qualitatively that they resemble the profiles shown in Figs.~\ref{fig2} (f) and (g). Then, the  ansatz~(\ref{eq:analyticalwf_3c}), which we proposed on the basis of the absence of restriction over the sign of the wavefunction when interchanging the third boson whit any of the other two, together with  ansatz~(\ref{eq:analyticalwf_3}) are giving valuable physical insight over the wavefunctions in this limit, and provide good 
seed functions to initiate calculations with numerical methods such as Diffusion Quantum Montecarlo (DMC)~\cite{Boronat:94}.

\begin{figure}
\begin{tabular}{cc}
\hspace{0.6cm}\includegraphics[width=4.1cm]{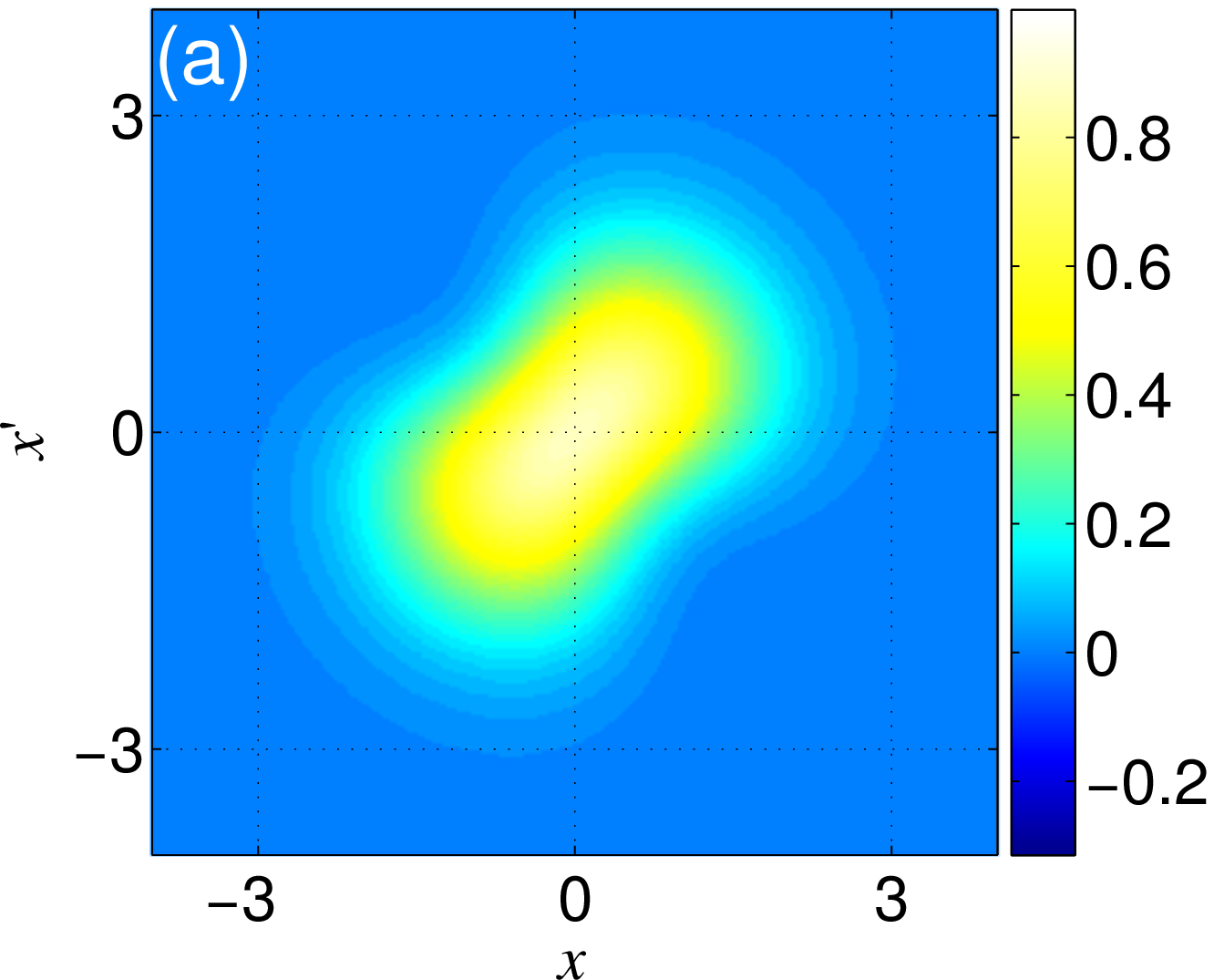}&\hspace{-0.1cm}\includegraphics[width=3.5cm]{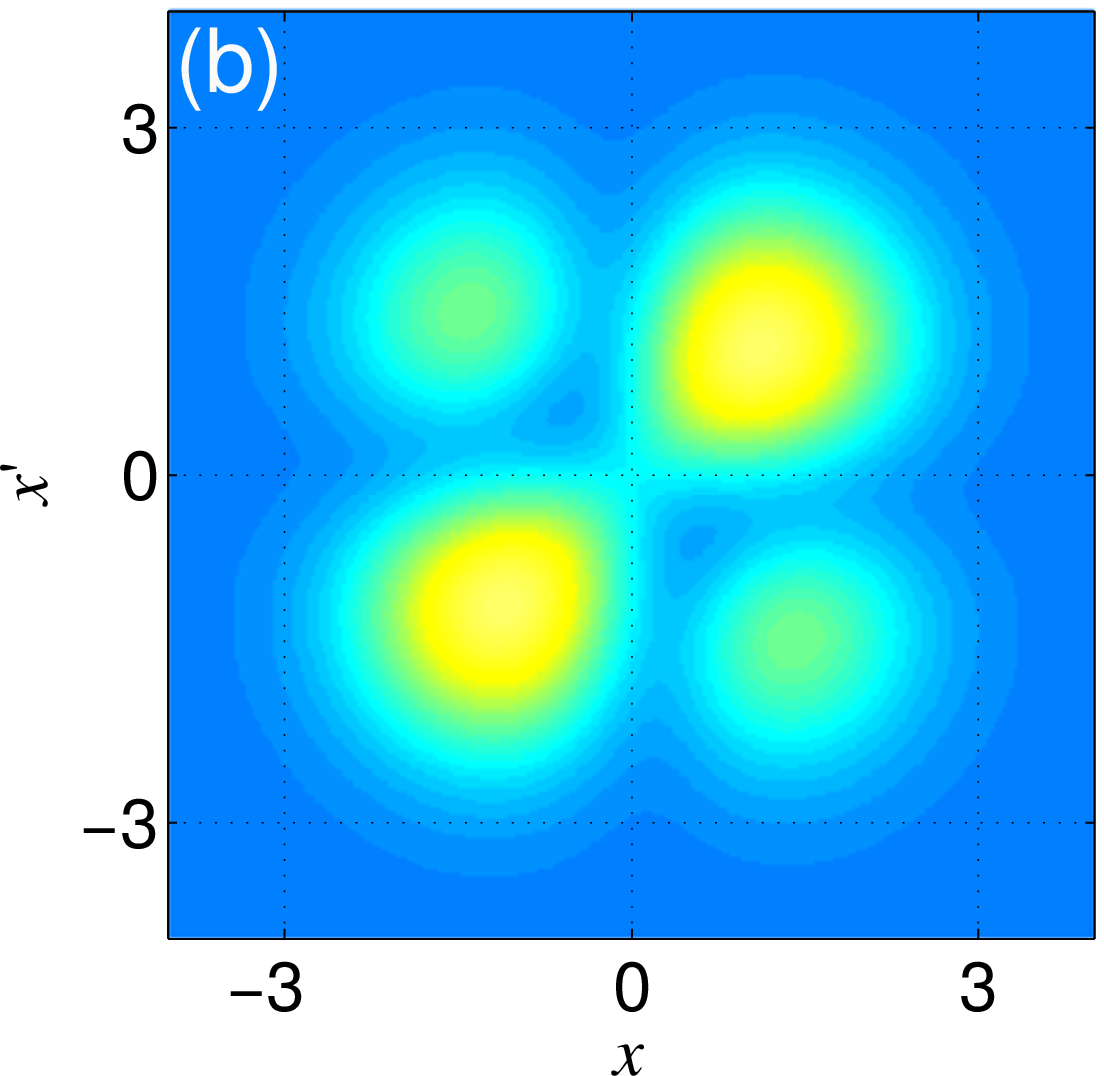}\\
\includegraphics[width=3.5cm]{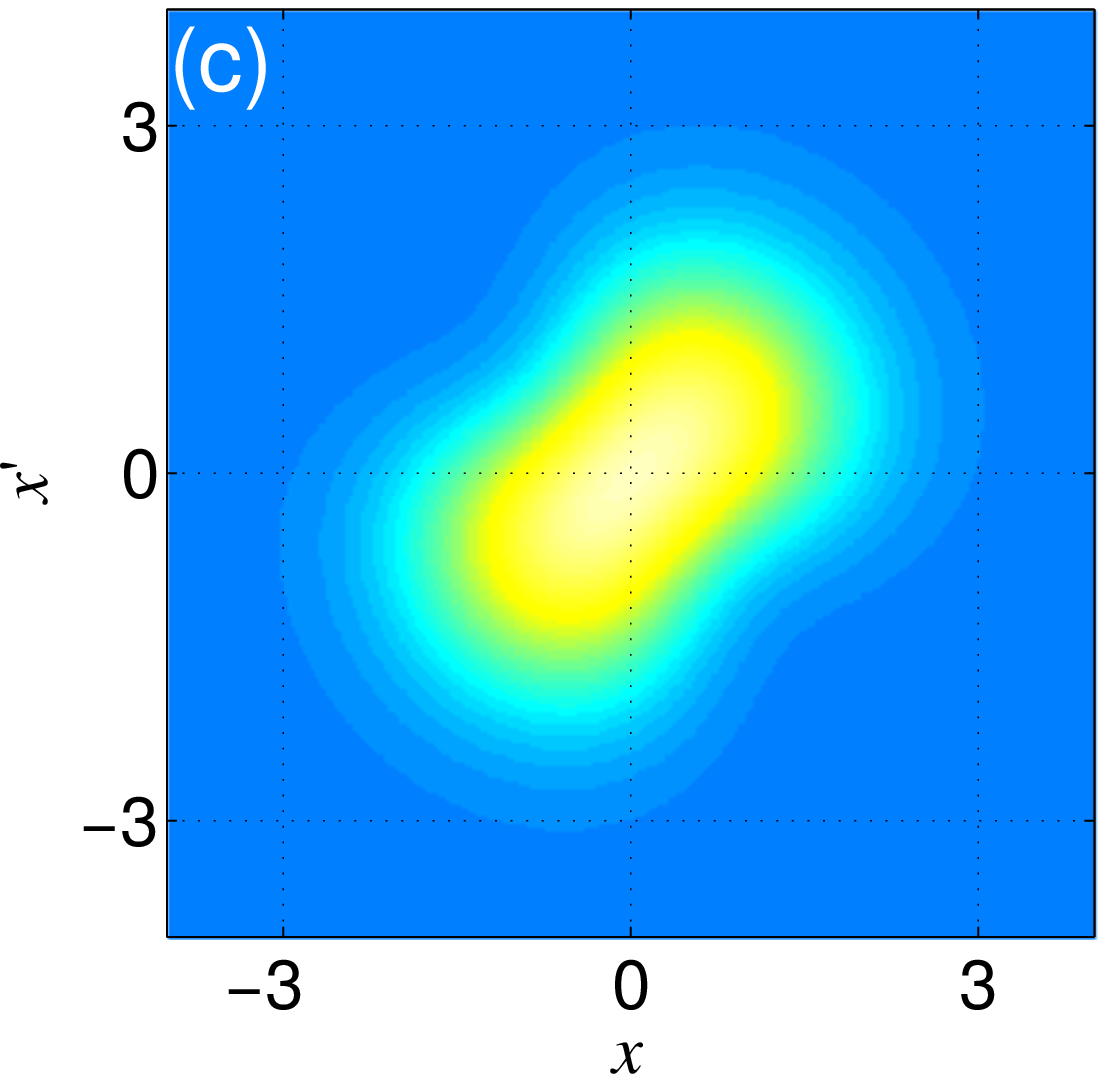}&\includegraphics[width=3.5cm]{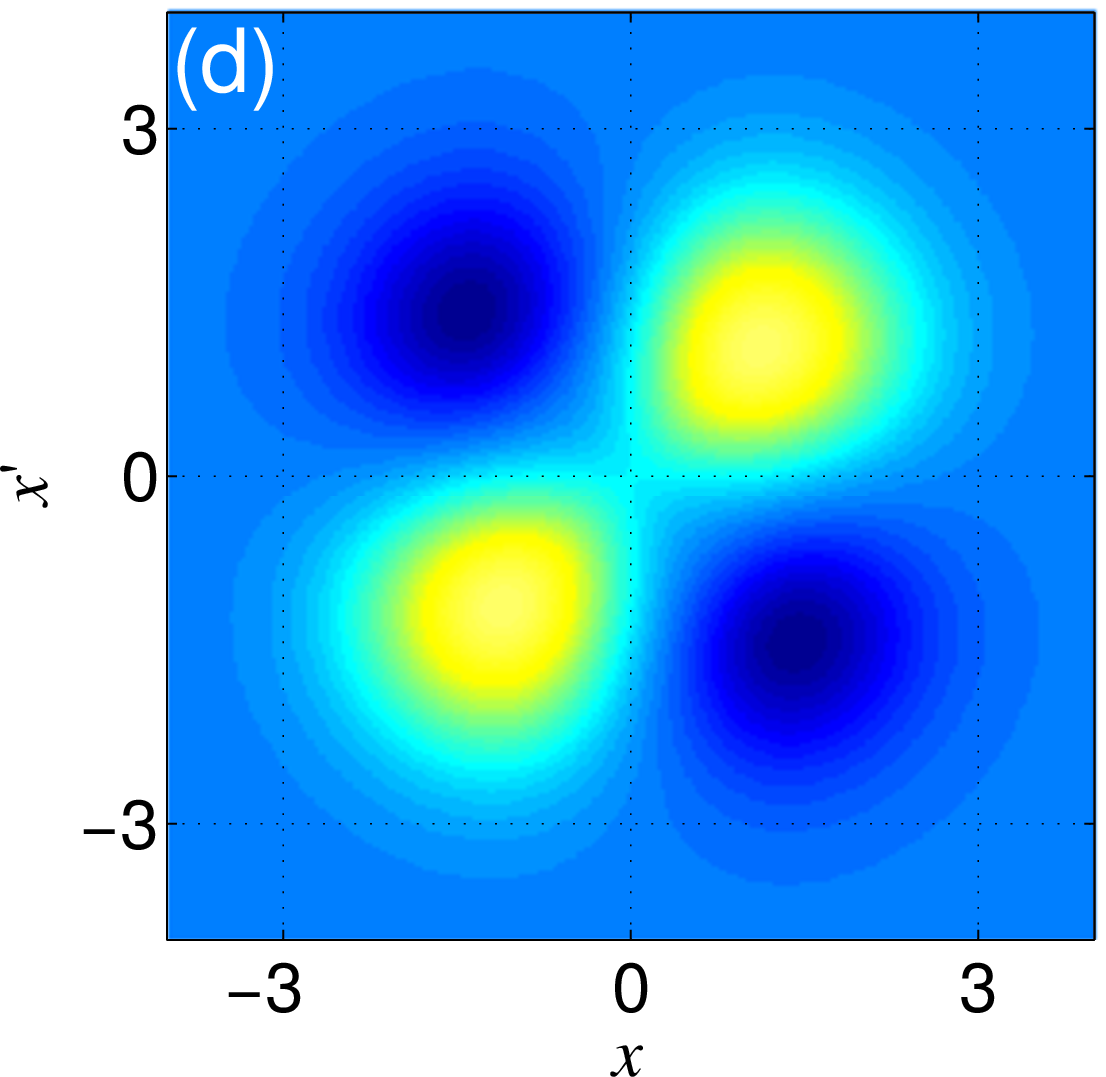}\\
\includegraphics[width=3.5cm]{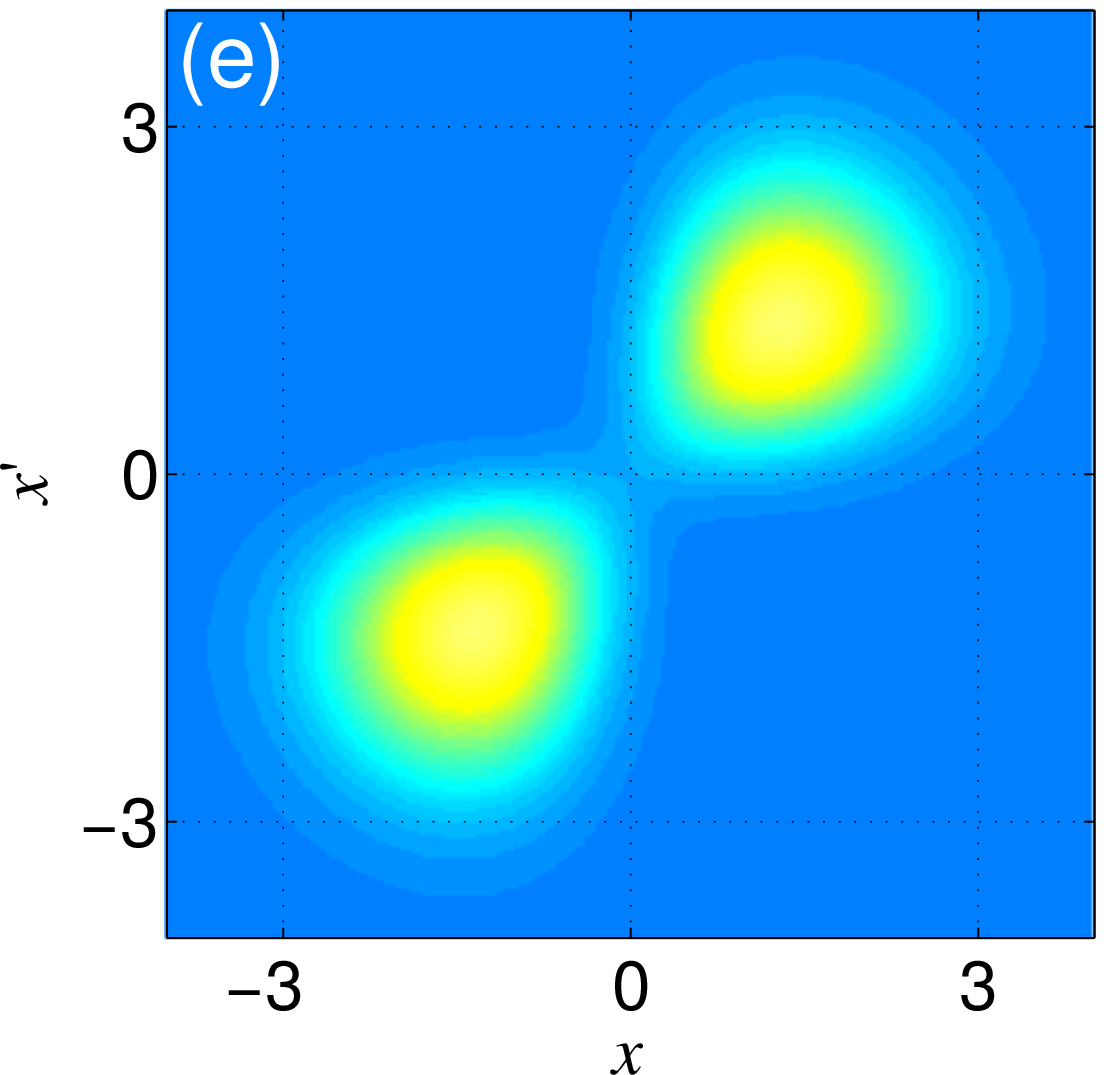}&\includegraphics[width=3.5cm]{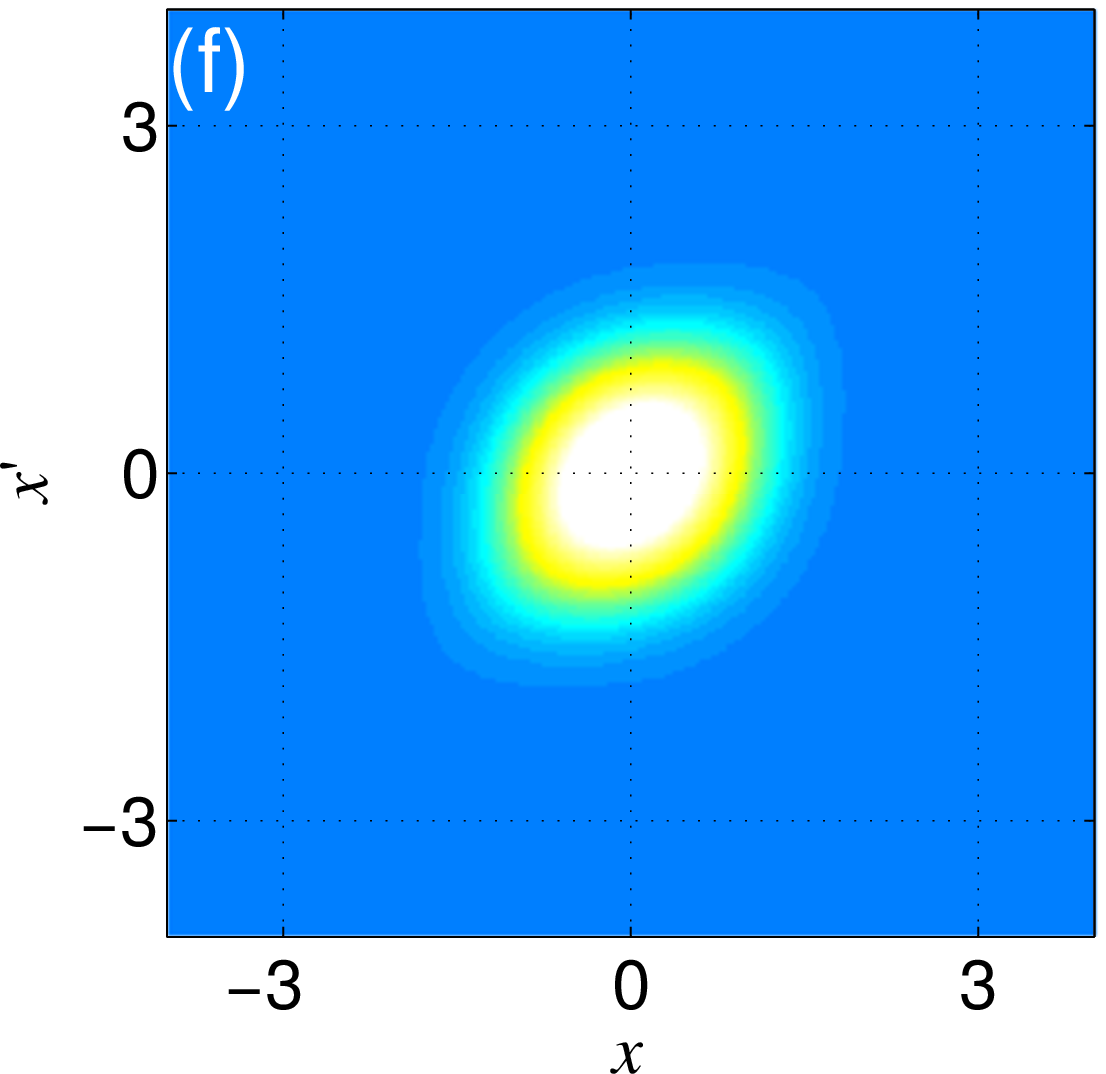}\\
\includegraphics[width=3.5cm]{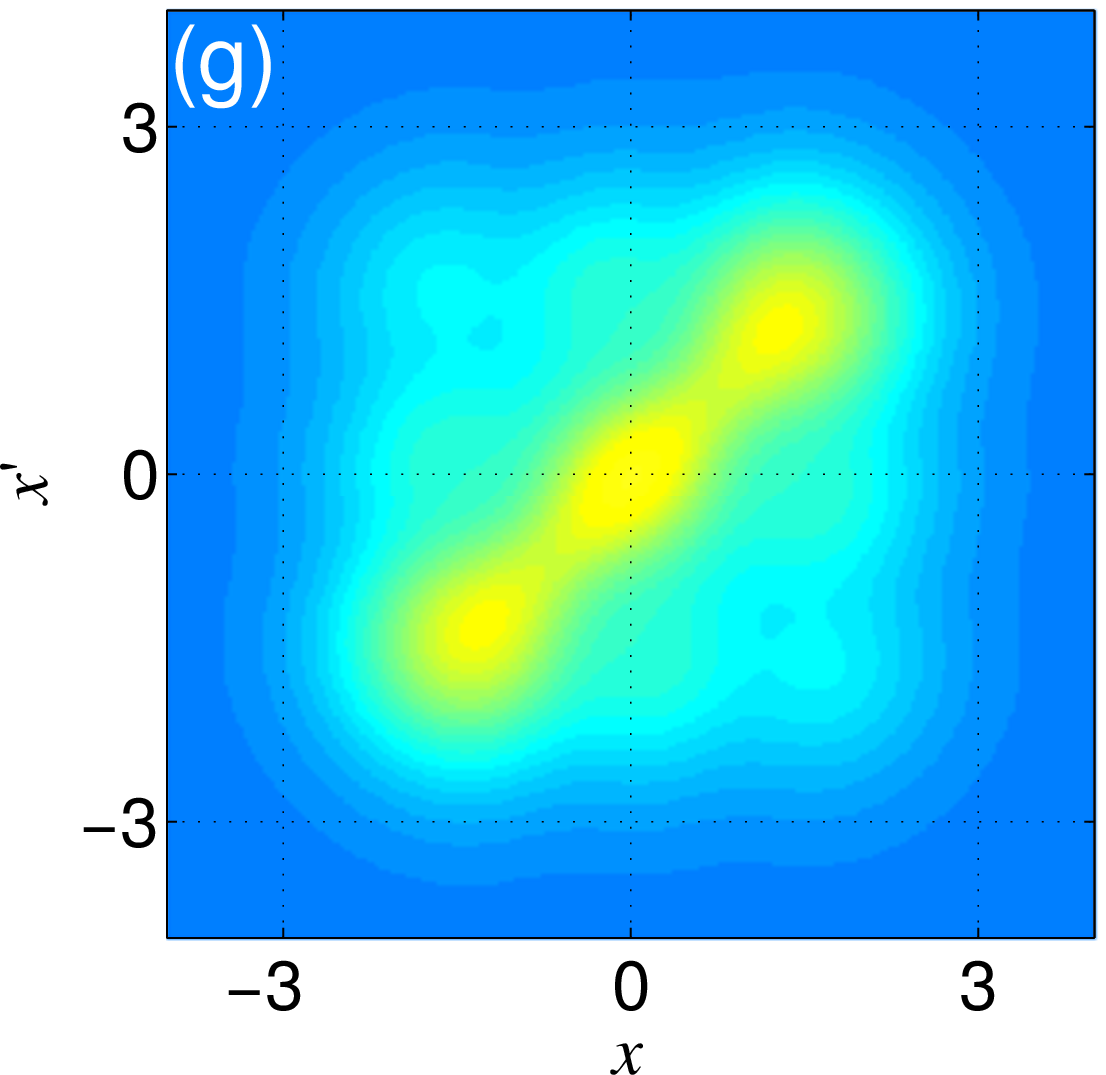}&\includegraphics[width=3.5cm]{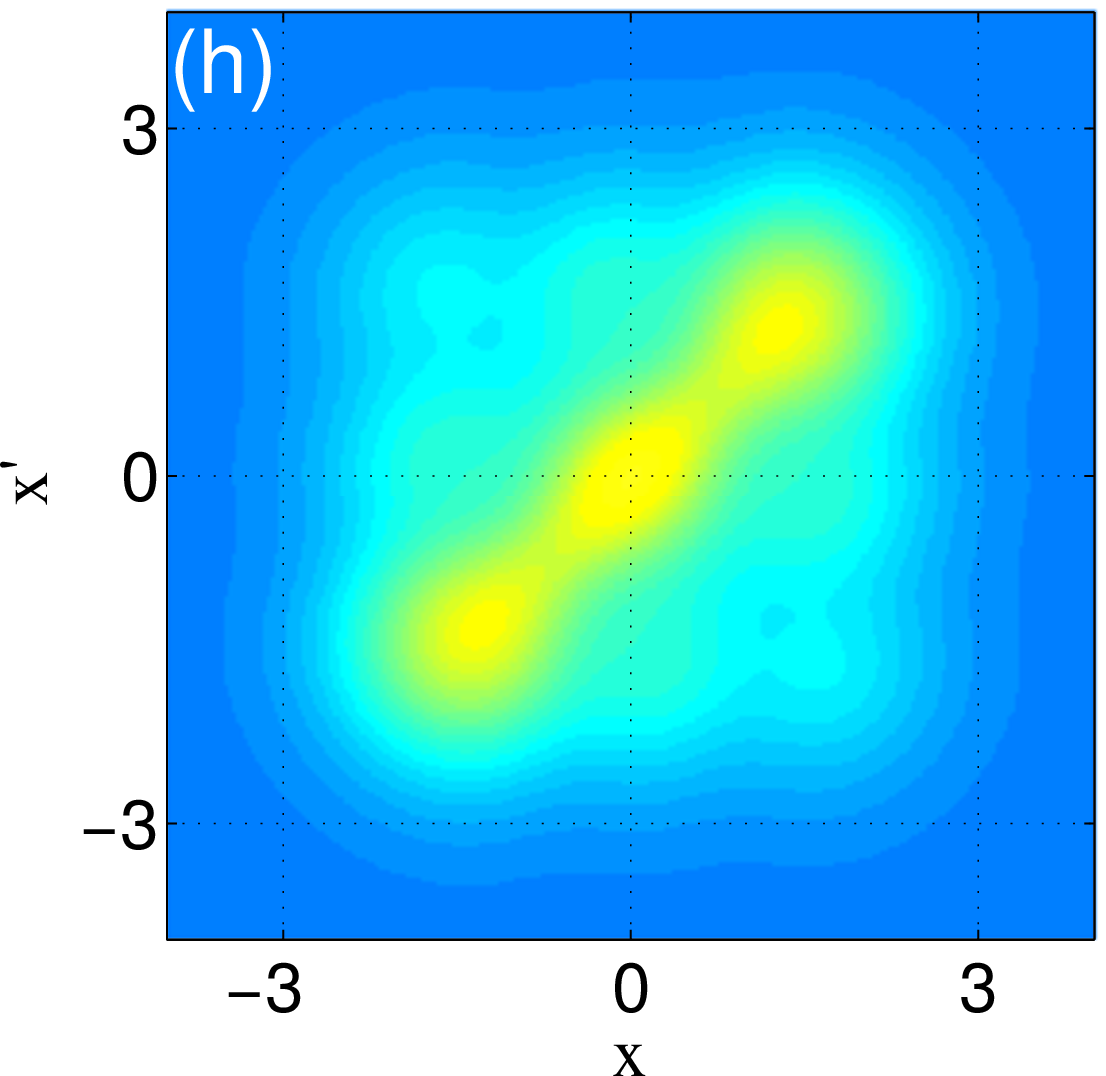}\\
\includegraphics[width=3.5cm]{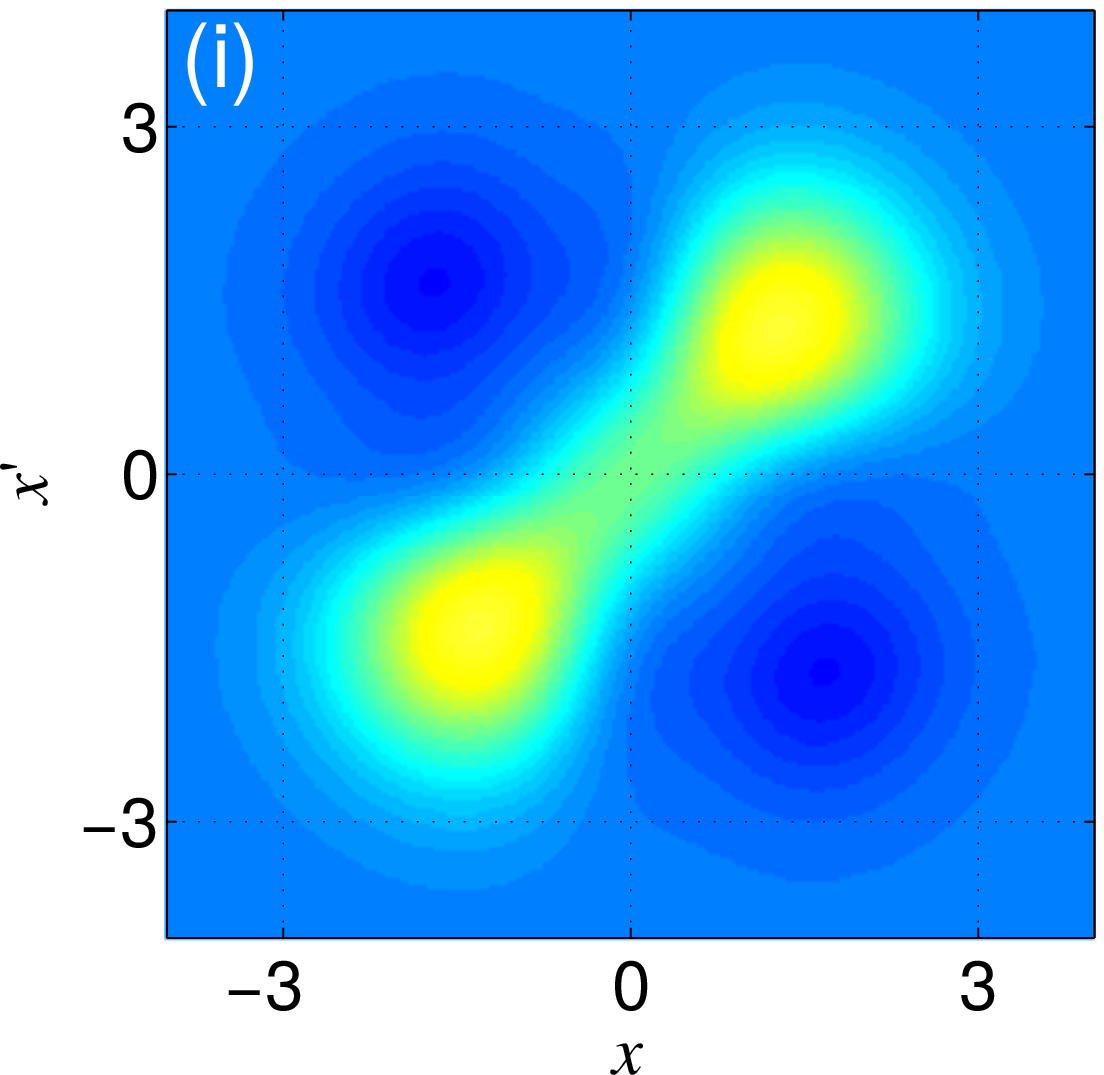}&\includegraphics[width=3.5cm]{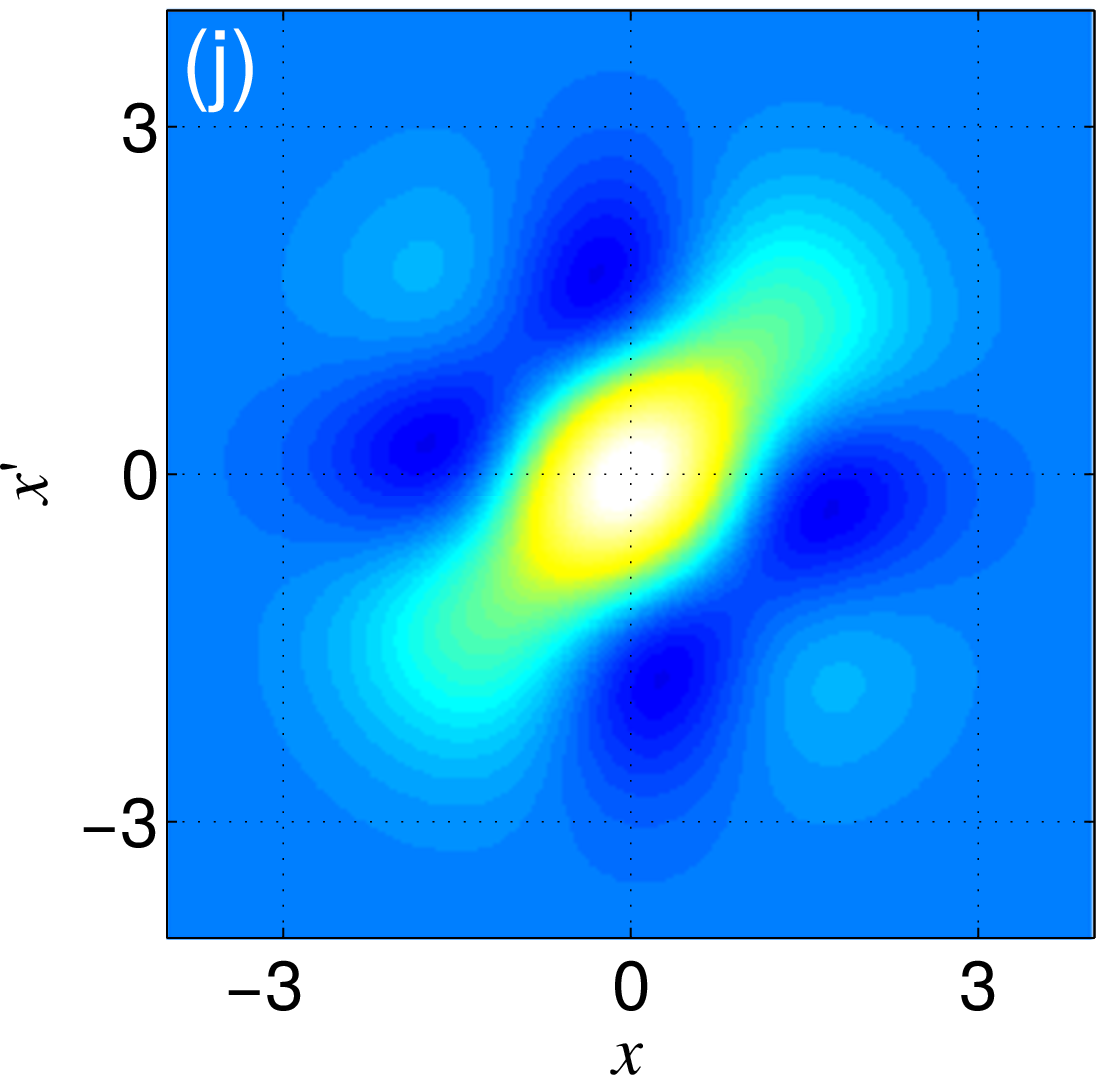}\\
\includegraphics[width=3.5cm]{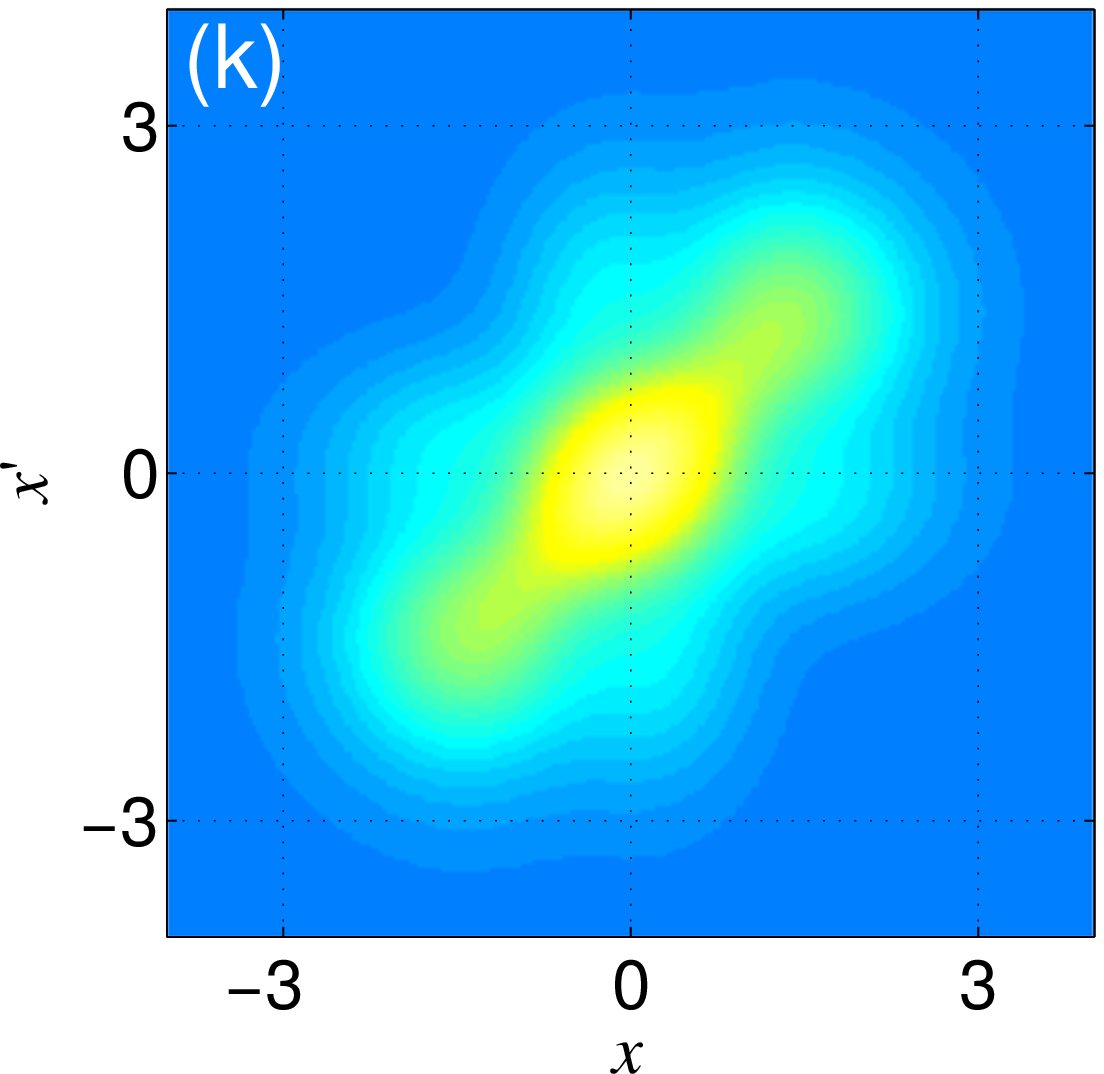}&\includegraphics[width=3.5cm]{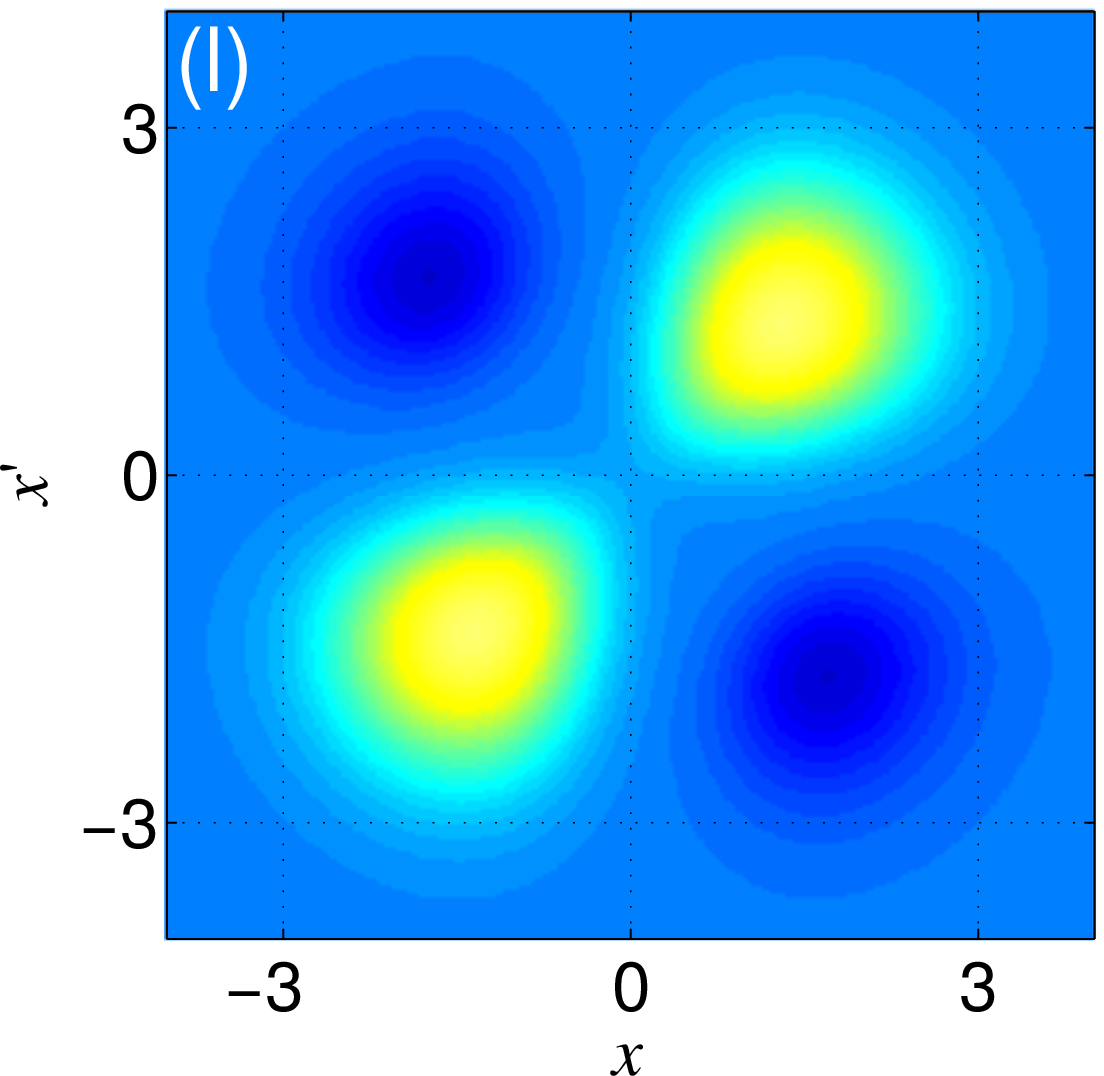}
\end{tabular}
\caption{(Color online)  One-body density matrices for A (left column) and B (right column). (a) to (f) correspond to the one-body density matrices calculated for the ground, first excited, and second excited states, when 
$g_{\mathrm{AB}}=50$ and $g_{\mathrm{A}}=0$. (g) to (l) to the ones calculated for the three degenerate states for   $g_{\mathrm{AB}}=g_{\mathrm{A}}=50. $\label{fig5}}
\end{figure}

 \subsection{Variable inter- and intra-species interactions}

 As the non-degenerate  second and seventh excited states obtained for large  $g_{\mathrm{AB}}$ and vanishing  $g_{\mathrm{A}}$  tend to zero  for $X=0$, their energy do not change greatly as $g_{\mathrm{A}}$ is increased, as observed in Fig.~\ref{fig1} (d). Indeed, for larger $g_{\mathrm{AB}}$, their energy remains constant as  $g_{\mathrm{A}}$ is changed, as shown in   Fig.~\ref{fig1} (e). 
For large $g_{\mathrm{AB}}$  we also observe that a three-fold quasi-degeneracy occurs for large $g_{\mathrm{A}}$ (see panel (e) of Fig.~\ref{fig1}). Similarly occurs for the spectra when both $g_{\mathrm{A}}$ and $g_{\mathrm{AB}}$ are kept equal while being  increased [see  Fig.~\ref{fig1}(f)]. In such a case, the degenerate states for $g_{\mathrm{A}}=g_{\mathrm{AB}}=0$ are the ones observed in the origin of Fig.~\ref{fig1}(a). Moreover, for  $g_{\mathrm{A}}=g_{\mathrm{AB}}$,  the interaction potential have the same strength  along the $Y$ axis and the lines $X=\pm\sqrt{3}Y$, and therefore the symmetry of the total interaction potential, Eq.~(\ref{eq:int_pot_total}), is always    $\mathcal{C}_{6v}$. Finally,  when  $g_{\mathrm{A}}=g_{\mathrm{AB}}$, the degeneracies occur only for $g_{\mathrm{A}}=0$ or for large $g_{\mathrm{AB}}$. These differences between the cases $g_{\mathrm{A}}=g_{\mathrm{AB}}$ and $g_{\mathrm{A}}\ne g_{\mathrm{AB}}$  have experimental implications: the ground state will differ if one 
follows a protocol  which increases both coupling constants at the same time or  a protocol which first  increases $g_{\mathrm{AB}}$ and then  $g_{\mathrm{A}}$. 
When  $g_{\mathrm{A}}$ and $g_{\mathrm{AB}}$ are large, only integer energy solutions remain~\cite{Zinner:13}. In the first column of Fig.~\ref{fig3} we represent the three wavefunctions when  $g_{\mathrm{A}} =g_{\mathrm{AB}}=2$.  In the second column of Fig.~\ref{fig3} we plot the three lowest energy quasi-degenerate wavefunctions when $g_{\mathrm{A}} =g_{\mathrm{AB}}=10$. These wavefunctions are degenerate when $g_{\mathrm{A}}$ and $g_{\mathrm{AB}}\to\infty$, and resemble the analytical prediction from Ref.~\cite{Zinner:13}.   The fact that the third atom is distinguishable is the responsible for the change of sign of the two excited state wavefunctions  when crossing the lines $X=\pm\sqrt{3}Y$, and therefore of the three-fold degeneracy occurring for infinite interactions. 
This has profound implications in the dynamics of the system, for example, after a quench in the AB interactions, where interesting phenomena as the emergence of orthogonality catastrophe has been found~\cite{Campbell:14}.

Finally, we find numerically that the first two wavefunctions [plotted in Fig.~\ref{fig3} (d) and (e)] have an overlap larger than $0.9$ with the ansatzs~(\ref{eq:analyticalwf_4b}) and~(\ref{eq:analyticalwf_4c}), respectively. The third one [plotted in Fig.~\ref{fig3} (f)] has also a similar overlap with ansatz~(\ref{eq:analyticalwf_4}), once this is orthonormalized to function~(\ref{eq:analyticalwf_4b}) by means of 
the Gram-Schmidt method. Again, these overlaps are approximate because they are numerically calculated. Moreover, the functions ~(\ref{eq:analyticalwf_4b}),~(\ref{eq:analyticalwf_4c}), and~(\ref{eq:analyticalwf_4}) orthonormalized to function~(\ref{eq:analyticalwf_4b}) resemble qualitatively the ones  plotted in the second column of Fig.~\ref{fig3}, when transformed to relative coordinates. Therefore, the functions~(\ref{eq:analyticalwf_4}) and~(\ref{eq:analyticalwf_4c}), which we proposed on the basis of the absence of a symmetrization condition over the distinguishable atom, turn to give physical insight on this limit, together with providing a good starting function for numerical methods such as DMC. 

\section{Coherence and  correlations}
\label{sec:coherence}

The interactions are responsible for building  up correlations and coherences  between the different atoms of a mixture of ultracold atoms  or  between the atoms within  the same species. In this Section we study the coherence and  correlations built up in the system along different protocols for varying the intra- and inter-species coupling constants. We also compare with those 
built up in the three identical boson problem and study  the possible spatial localization patterns in the  different limits. 

\subsection{Correlations along different protocols for varying the interactions and comparison with three identical bosons}

Let us first find to which extend the mixture of two bosons with a third, distinguishable one, resembles a TG gas of three atoms only for certain values of the coupling constants. 
To quantify the coherence in a system of three identical bosons one can study the one-body density matrix (OBDM)  which,  in second quantization, is $\rho_{(1)}(x,x')=\sum_{k,k'}\phi_{k}(x)\phi_{k'}(x')\left\langle
\right.\!\! a_{k}^{\dagger}a_{k'}\!\!\left.\right\rangle $, where  $\phi_{k}(x)$ are the harmonic oscillator modes  used in the expansion (see Sec.~\ref{sec:numerical}).  Diagonalization of the OBDM produces the natural orbitals and their occupations, $\lambda_i$. The largest occupation of a natural orbital, $\lambda_0$, provides information about the degree of Bose-Einstein condensation in the system, and therefore of the coherence between the atoms. We plot in Fig.~\ref{fig4}~(a) the spectra of eigenenergies for three indistinguishable atoms and in (b) the  largest occupation of a natural orbital, both as a function of the coupling constant $g$.  As $g$ is increased from zero to a large value, the energy increases from   $E=1.5$ (three ideal bosons) to $E=4.5$ (three TG bosons). Also, the largest occupation is reduced from 1  to its corresponding value for a TG gas, $\lambda^0\simeq N^{-0.41}=0.63$~\cite{Girardeau:01}.

There are different possibilities of building   correlations in a system of two A atoms and a third B atom, as there are two coupling constants, i.e., $g_\mathrm{A}$ and $g_\mathrm{AB}$. If we increase the coupling constants following a protocol that keeps both of them equal, $g_\mathrm{A}=g_\mathrm{AB}$, the obtained ground state is similar to that of a system of three indistinguishable atoms. In Fig.~\ref{fig4} (a) we plot both  the energy of the ground state of three indistinguishable atoms and that obtained for the ground state as $g_\mathrm{A}=g_\mathrm{AB}$ is increased for two A atoms and one B atom [which corresponds to solid line in Fig.~\ref{fig1} (f)]. Both curves coincide. In the system of two A atoms and one B atom, the OBDM can be calculated either for an A atom or the B atom, $\rho_{(1)}^{\mathrm{A}}(x,x')=\sum_{k,k'}\phi_{k}(x)\phi_{k'}(x')\left\langle
\right.\!\! a_{k}^{\dagger}a_{k'}\!\!\left.\right\rangle $ and $\rho_{(1)}^{\mathrm{B}}(x,x')=\sum_{k,k'}\phi_{k}(x)\phi_{k'}(x')\left\langle
\right.\!\! b_{k}^{\dagger}b_{k'}\!\!\left.\right\rangle $, respectively. We diagonalize both matrices for increasing values of the coupling constants, keeping $g_\mathrm{A}=g_\mathrm{AB}=g$, and plot the corresponding largest occupation of a natural 
orbital in each case, $\lambda_0^{\mathrm{A(B)}}$, in Fig.~\ref{fig4} (b). Note that the occupations for A and B coincide as $g$ is varied. Furthermore,  they also  coincide with the one obtained for three indistinguishable atoms. 

One can also calculate  the von Neumann entropy (vNE)  associated to the OBDM, defined as  $S(\rho_{(1)})=-\text{Tr}[\rho_{(1)}~\text{log}_2~\rho_{(1)}]$, which can be obtained from the natural orbits occupation as
$ S(\rho_{(1)})=-\sum_i \lambda_i \text{log}_2 \lambda_i\;$. We plot the vNE as a function of $g$ for the system of three indistinguishable atoms in Fig.~\ref{fig4} (b). For two A atoms and a third B atom, one calculate the vNE from the OBDM of B, thus corresponding to tracing out all degrees of freedom of the two A atoms.  In Fig.~\ref{fig4} (b) we plot the vNE of the system of two A atoms and a third B atom when it is calculated from the OBDM of B. This vNE coincides with the one of the system of three indistinguishable atoms, as $g_\mathrm{A}=g_\mathrm{AB}$ is increased. We remark that when $g_\mathrm{A}=g_\mathrm{AB}$ the  symmetry of the total interaction potential, Eq.~(\ref{eq:int_pot_total}), is always    $\mathcal{C}_{6v}$ and that the ground-state wavefunction is positive and fulfills this symmetry [see Fig.~\ref{fig3} (a) 
and (d)]. 
We conclude that the transition from the non-interacting gas of two A atoms and a single B atom when both  $g_\mathrm{A}$ and $g_\mathrm{AB}$ are increased while kept equal is analogous to the one for three indistinguishable atoms. Indeed the ground state is  non-degenerate for a large interval of values of  $g_\mathrm{A}=g_\mathrm{AB}$. For example, the state in Fig.~\ref{fig3} (d) is very similar to the wavefunction of a TG gas of three atoms. Indeed, as we already discussed,  it has an overlap larger than $0.9$ with ansatz~(\ref{eq:analyticalwf_4b}), which is the analytical wavefunction of a three-atom TG gas. However, one important difference is  that there exists a three-fold degeneracy for   $g_\mathrm{A}=g_\mathrm{AB}\to\infty$ [see Fig.~\ref{fig1} (f)] which does not occur in the case of three identical bosons.  In short, while for finite interactions the  mixture is similar  to a gas of three atoms,  for infinite interactions one has to take into account that there are two more states degenerate in 
energy 
with this one. 

The system of two A atoms and a single B atom permits one to build  correlations by following different protocols than that resembling a three atom system, i.e., keeping $g_\mathrm{A}=g_\mathrm{AB}$. Another possible protocol to increase  both $g_\mathrm{A}$ and $g_\mathrm{AB}$ towards infinity, consists in increasing  $g_\mathrm{AB}$ keeping $g_\mathrm{A}=0$ in a first stage, and then increasing   $g_\mathrm{A}$ in a second stage.  Throughout the first stage the interaction potential is  given by Eq.~(\ref{eq:int_pot_AB}), which shows symmetry $\mathcal{C}_{2v}$. Then, this two-stage protocol does not allow for solutions similar to the ones obtained for the three indistinguishable atom case.  In Fig.~\ref{fig4} (c) we show the largest natural orbital occupation for A and B for the ground state as $g_\mathrm{AB}$  is increased, with $g_\mathrm{A}=0$. As shown, the largest natural orbit occupation is different for each species, and different from the three atom case. Also, the vNE calculated from the OBDM of B 
increases 
from zero to a value of ~1.2 for  large $g_\mathrm{AB}$, which is smaller than the one reached when using the equal-coupling-constant protocol. For $g_\mathrm{AB}$ large with $g_\mathrm{A}=0$, we reach the limit in which the ground state and first excited states are quasi-degenerate  [see Figs.~\ref{fig1} (d) and Figs.~\ref{fig2} (f) and (g)]. In Fig.~\ref{fig4} (d) we plot the largest natural orbit occupation for A and B and the vNE along the second stage,  starting from  the even-parity ground state obtained for  $g_\mathrm{AB}=50$ and 
$g_\mathrm{A}=0$. Note that we are adding to the interaction potential given by Eq.~(\ref{eq:int_pot_AB})   the one given by Eq.~(\ref{eq:int_pot_A}), but still the symmetry of the total interaction potential is  $\mathcal{C}_{2v}$. 
We observe that for large values of $g_\mathrm{A}$ the occupations are different and smaller than the ones obtained in the large $g_\mathrm{A}$ and $g_\mathrm{AB}$ limit  when using the equal-coupling-constant protocol. Also, the vNE reached in this limit is smaller when using the two-stage protocol, than that reached when using the equal-coupling-constant protocol. Note that, for large and equal  coupling constants we have restored the $\mathcal{C}_{6v}$ symmetric potential  given by Eq.~(\ref{eq:int_pot_total}), and the system shows a three fold quasi-degeneracy.

\subsection{Spatial localization and degeneracy}

The OBDMs contain information about the spatial localization 
of the two A atoms and the B atom. Indeed, their diagonal $x=x'$  for the A atoms (or $y=y'$ for the B atom) provides the actual density profile. 
  
Let us first analyze OBDMs of the ground and first two excited states in the limit of large  $g_\mathrm{AB}$ with $g_\mathrm{A}=0$ [plotted in Fig.~\ref{fig5} (a) to (f)]. 
The OBDMs in Figs.~\ref{fig5} (a) and (b) [which corresponds to the  ground state shown in Fig.~\ref{fig2} (f) with ansatz given by Eq.~(\ref{eq:analyticalwf_3})] show that the two A atoms tend to stay localized in the center of the trap while the single B atom stays at the edges. Similarly occurs for the first excited state, which is quasi-degenerate with the ground state in this limit  [see Figs.~\ref{fig5} (c) and (d)]. This state is the one  shown in Fig.~\ref{fig2} (g)  with ansatz Eq.~(\ref{eq:analyticalwf_3c}). 
This configuration can be also observed in   Figs.~\ref{fig2} (f) and (g) because 
the  values of the wavefunction significantly differ from zero when the relative coordinates are located around the $Y$ axis and between the lines  $X=\pm\sqrt{3}Y$, which means that both $x_1$ and $x_2$ are smaller than $y$. The fact that for these two quasi-degenerate states there is a great overlap between the density profiles of A and B is  reminiscent of the composite fermionization process described in this limit for larger number of atoms of B~\cite{Zollner:08a,Garcia-March:13, Garcia-March:14}. 

Contrarily, from the OBDMs of the second excited state [Figs.~\ref{fig5} (e) and (f)]  one recognizes that the atom  B tends to stay localized in the center of the trap, while the two A atoms spatially separate. This is also in accordance with Fig.~\ref{fig2} (h), as now the significantly non-zero values of the wavefunction in relative coordinates are  located around  the $X$ axis and between the lines  $X=\pm\sqrt{3}Y$, which corresponds to $x_1$ and $x_2$ larger than $y$. 

The spatial localization patterns are different in  the limit  $g_\mathrm{A}$ and  $g_\mathrm{AB}$ large, for the ground and first two excited states, which in this case are quasi-degenerate (see  Figs.~\ref{fig5} (g) to (l)).  
 The OBDMs  plotted in Figs.~\ref{fig5} (g) and (h) [which correspond to the state plotted in Fig.~\ref{fig3} (d) with  ansatz given by Eq.~(\ref{eq:analyticalwf_4b})] are similar to that of a  TG gas of three atoms, and therefore both OBDMs are equal. This is the ground state  reached if $g_\mathrm{AB}$ and $g_\mathrm{A}$ are increased at the same time towards infinity. 
 The OBDMs plotted in Figs.~\ref{fig5} (i) and (j) [corresponding to the state plotted in  Fig.~\ref{fig3} (f) with ansatz Eq.~(\ref{eq:analyticalwf_4})] show that, for this state,  the B atom stays in the center of the trap, while the A atoms stay in the edges. Finally, the OBDMs plotted in Figs.~\ref{fig5} (k) and (l) [which correspond to the state shown in  Fig.~\ref{fig3} (e) with the ansatz of Eq.~(\ref{eq:analyticalwf_4c})] illustrate that, for this state, the B atom is mainly spatially located in 
the edges of the trap, while the A atoms stay in the center. In short, the spatial localization of the A and B atoms when some coupling constant is large will depend on the protocol followed to reach this limit, as it determines the actual lowest energy wavefunction which will be actually reached.

\begin{figure}
\includegraphics[width=0.95\columnwidth]{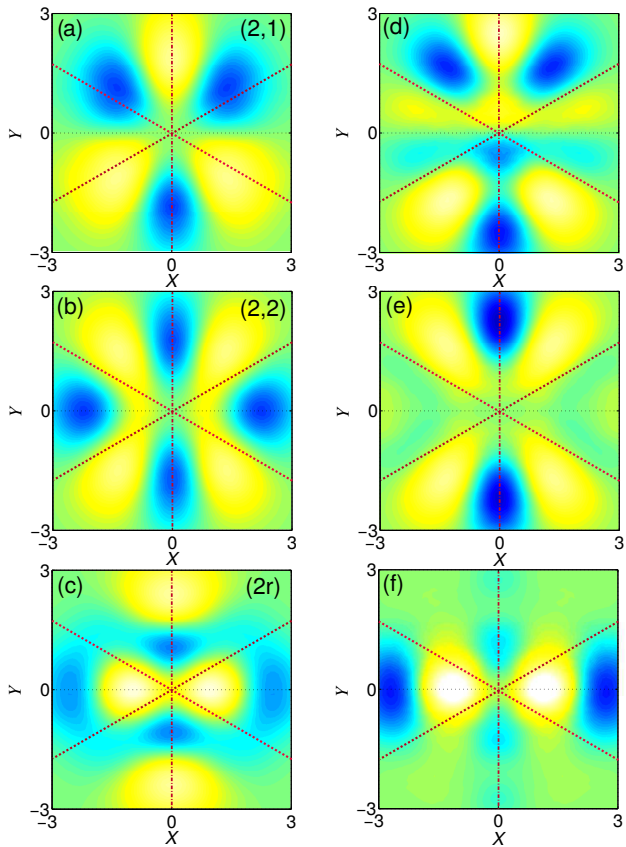}
\caption{(Color online) {\it  Wavefunctions of the fifth to seventh excited states in the $X$-$Y$ plane, when  $g_{\mathrm{A}} =0$ and $g_{\mathrm{AB}} =2$ and 10}. (a) and (b) represent the fifth and sixth quasi-degenerate excited states when $g_{\mathrm{AB}} =2$. (c) represents  non-degenerate seventh excited state for $g_{\mathrm{AB}} =2$. (d)  to (f)  are the fifth to seventh state when $g_{\mathrm{AB}} =10$. Dashed (dash-dotted) red  lines highlight the axis along the which the AB (A) interactions occur. In panels (a) to (c), we indicate the number of nodes in the $X$ ($N_X$)  and $Y$ ($N_Y$) directions  as $(N_X,N_Y)$. Instead, when $N$ radial nodes occur, we indicate them as $(Nr)$.  \label{fig7}}
\end{figure}

\section{Conclusions}
\label{Sec:conc}

We present a comprehensive study of the ground and  excited states of a mixture of two identical bosons and a third distinguishable atom in a one-dimensional parabolic trap. In this system, there are two different types of  interactions, that is, the intra-species interactions between the two identical bosons and the  inter-species interactions between these two and the third distinguishable atom. We assume contact, repulsive intra- and inter-species interactions governed by the coupling constants $g_\mathrm{A}$ and $g_\mathrm{AB}$, respectively.  By  writing the system's Hamiltonian  in center-of-mass and relative coordinates, we find the locus of the points where the intra- and inter-species interactions take place, which occur along certain axis in the relative coordinate plane. We find that the distinguishability of the third atom means that there is no restriction over the  sign of the wavefunction when crossing the axis  associated with the inter-species interactions. On the contrary, the 
wavefunction has to be always positive when crossing the axis associated with the intra-species interactions.  The absence of such a restriction is at the origin of the presence of degeneracies when the coupling constants reach some limiting cases, that is, when any or both of them  tend to infinity. We  propose physically meaningful ansatzs based on the expected fermionization of the two identical  bosons when the intra-species interactions tend to infinity and in a similar behavior when the inter-species interactions tend to infinity. In the latter case, since there is no need to restrict the wavefunction to be positive, we find that more than one ansatz is possible, which is explained by the presence of degeneracies. The  interaction potential is a sum of delta functions in the plane where the relative motion occurs, with a strength $g_\mathrm{A}$ or $g_\mathrm{AB}$~\cite{Harshman:12}. If the strengths of both 
interactions is equal, this potential is $\mathcal{C}_{6v}$ discrete rotationally symmetric, while if they are different, it shows $\mathcal{C}_{2v}$ discrete symmetry.  This allows us to use discrete group theory to analyze and extract restrictions over the form of the ground and excited states wavefunction dictated by this symmetry, and further understand the degeneracies observed when  any or both coupling constants tend to infinity.  

We first observed that the degeneracies that exist when both $g_\mathrm{A}$ and $g_\mathrm{AB}$ vanish, are broken for small and finite $g_\mathrm{A}$, with vanishing  $g_\mathrm{AB}$. We found that, for  $g_\mathrm{A}=0$,  the ground and excited states are well described by products of the one-dimensional eigenfunctions  of the harmonic oscillator. These can be written in terms of Hermite polynomials and labeled by two quantum numbers, $n_X$ and $n_Y$. The symmetrization condition over the two identical particles imposes that $n_X$ has to be even, which permitted us to identify the ground state and all possible excitations at  $g_\mathrm{A}=0$. For finite and small  $g_\mathrm{A}$, these two quantum numbers still characterize well the solutions, and particularly determine the number of nodes in the $X$ and $Y$ directions. We used the spectra for   $g_\mathrm{AB}=0$ and varying $g_\mathrm{A}$ as a benchmark to study the spectra when $g_\mathrm{AB}\ne0$. We found that the ground state is two-fold degenerate 
when $g_\mathrm{AB}\to\infty$ and $g_\mathrm{A}=0$, with a structure of excitations that included both double-degenerated states and singlets. This is in accordance with the results 
from Ref.~\cite{Zinner:13}.   We further found that the ground state is three-fold degenerate when both $g_\mathrm{A}$  and $g_\mathrm{AB}\to\infty$, with excited states which are also three-fold degenerate (again in accordance with Ref.~\cite{Zinner:13}). We compared  the numerically calculated states in both limits with the aforementioned ansatzs,  finding very good agreement. 

We studied how  coherence and  correlations evolve in the ground and excited states as the interactions are tuned. Since there are two independent  coupling constants in the mixture, one can tune the interactions following different protocols. We showed that along an equal-coupling-constant protocol, that is, tuning both coupling constants and keeping them always equal, the ground state of the system corresponds to that of a three identical boson system for the whole range of interactions. Importantly, even for the equal-coupling-constant protocol, in the infinite coupling constant limit the system is three-fold degenerate, which is a crucial difference between this mixture and the three-boson system.  On the contrary, we found that through two-stage protocols, that is, increasing first one coupling constant towards infinity and then the second one, the ground state is different from that of the three identical boson system. We noted that, on the one hand, the  different possible ground states 
correspond to different spatial distributions of the two identical bosons with respect to the third distinguishable atom. On the other hand,  it also corresponds to different  correlations built up between the two species and to a different degree of  condensation in each species. Finally, since  there exist degeneracies in the infinite coupling constant limits, the actual protocol followed to reach this limit is crucial to determine the  ground state of the system realized in practice.

\section{Acknowledgments}
    We acknowledge  partial financial support from the
  DGI (Spain) Grant No. FIS2011-25275, FIS2011-24154 and  the Generalitat de Catalunya Grant No. 2014SGR-403. GEA and BJD are supported by the Ram\'on y Cajal program, MEC (Spain).
  The authors acknowledge useful discussions with Thomas Busch.

  \appendix
  
\section{Direct diagonalization method}
\label{sec:numericalmethod}

 To numerically  calculate the properties of the ground and excited states of the systems, we employ the second-quantized exact diagonalization algorithm described in~\cite{Garcia-march:12},
 which makes use of the expansion of the second-quantized field operator of A in terms of many modes, that is, 
\begin{equation}
 \label{eq:expansion}
\hat \Psi^{\mathrm{A}}(x)=\sum_{k}a_{k}\phi_{k}(x) , 
\end{equation}
where  the modes $\phi_{k}(x)$ are the eigenfunctions of the single-particle part of the Hamiltonian, $H_{\mathrm{sp}}=- \frac {1}{2} \frac {d^2}{dx^2} + \frac {1}{2} x^2$ and the creation/annihilation operators $a_{k}$ and $a_{k}^\dagger$ satisfy bosonic commutation relations. We use the same expansion for the field operator of B, $\hat \Psi^{\mathrm{B}}(x)$, by introducing the corresponding bosonic  creation/annihilation operators $b_{k}$ and $b_{k}^\dagger$. With this procedure we derive a  many-mode Hamiltonian which we express in terms of a Fock basis 
\begin{equation}
\label{eq:FockBasis}
 \Phi_i=D_i\left(\hat{a}_{1}^\dagger\right)^{N_{1,i}^\mathrm{A}}\!\!\!\!\dots
         \left(\hat{a}_{n_\mathrm{A}}^\dagger\right)^{N_{n_\mathrm{A},i}^\mathrm{A}}
         \left(\hat{b}_{1}^\dagger\right)^{N_{1,i}^\mathrm{B}}\!\!\!\!\dots
         \left(\hat{b}_{n_{B}}^\dagger\right)^{N_{n_\mathrm{B},i}^\mathrm{B}}\Phi_0,
\end{equation}
where $D_i=(N_{1,i}^\mathrm{A}!\dots N_{n_\mathrm{A},i}^\mathrm{A}!)^{-\frac{1}{2}}$ and $\Phi_{0}$ is the vacuum. The occupation numbers of the $n_\mathrm{A}$ ($n_{B}$) modes for each component are given by  $N_{1,i}^\mathrm{A},\dots,N_{n_\mathrm{A},i}^\mathrm{A}$ ($N_{1,i}^\mathrm{B},\dots,N_{n_{B},i}^\mathrm{B}$). Note that they can only take values from 0 to 2 for A, and 0 or 1 for B.  The dimension of the Hilbert space  is
$\Omega=\Omega_\mathrm{A}\Omega_{B}$ with
$\Omega_\mathrm{A}=(n_\mathrm{A}+1)n_\mathrm{A}/2$ and 
$\Omega_\mathrm{B}=n_\mathrm{B}$. After diagonalizing this Hamiltonian we obtain the eigenenergies and the ground and excited states, which we express as expansions in terms of the  of Fock vectors $\Psi_{j}=\sum_{i=1}^{\Omega}c_{i}^j\Phi_{i}$. To obtain  the wavefunction in first quantization from  the numerically calculated one in second quantization by direct diagonalization, we consider all possible permutations of the two indistinguishable atoms over the single-particle basis used in the many-mode expansion of the field operator, Eq.~(\ref{eq:expansion}).

  \section{Highly excited states when varying $g_\mathrm{AB}$ with $g_\mathrm{A}=0$}
  \label{sec:appendixB}
  
  In Fig.~\ref{fig7} we show the  fifth to seventh excited states when $g_\mathrm{AB}=2$ (left column) and $g_\mathrm{AB}=10$ (right column), and $g_\mathrm{A}=0$. The corresponding ground and first four excited states were shown in Fig.~\ref{fig2}. The third  and fourth excited states  are a  radial excitation of the ground and first excited states along the $Y$ axis. As there is no symmetry restriction along this axis, excitations with an odd number of nodes in this   direction are permitted. The fifth and sixth excited states are a second excitation of the ground and first excited states, now along a horizontal direction. There is a restriction on the number of nodes in this direction, as there can not be changes of sign in $X=0$, and therefore only even excitations are permitted. The seventh excited  state is a radial excitation of the second excited state, which now can show again  both an even or an odd number of nodes, as the nodes do not coincide with the $Y$ axis.

\end{document}